\documentclass[aps,12pt]{revtex4-2}

\usepackage{graphicx}
\usepackage{dcolumn}
\usepackage{amsmath}
\usepackage{amssymb}
\usepackage{amsbsy}
\usepackage{float}
\usepackage{url}
\usepackage[bookmarksopen]{hyperref}
\usepackage{caption}
\usepackage{physics}
\usepackage{float}
\usepackage{natbib}

\begin{document}

\title{Breaking QAOA's Fixed Target Hamiltonian Barrier: A Fully Connected Quantum Boltzmann Machine via Bilevel Optimization}

\author{Liu Jun}
\email{liujun@hufe.edu.cn}
\email{lewjhon@126.com}
\affiliation{School of Economics, Hunan University of Finance and Economics, Changsha 410205, China}

\begin{abstract}
	\textbf{Abstract} To overcome the limitations of classical partially connected Boltzmann machines and mainstream quantum Boltzmann machines (QBMs), this work extends the conventional circuit of the quantum approximate optimization algorithm (QAOA) to a bilevel optimization architecture and proposes a fully connected QBM. Specifically, this work breaks the constraint that the target Hamiltonian in standard QAOA circuits is fixed by setting its structural parameters as optimizable variables, which are used to map the energy function parameters of the fully connected Boltzmann machine. The inner-loop training simulates positive phase energy minimization based on the computational process of the conventional QAOA circuit, whereas the outer-loop training simulates negative phase contrastive divergence learning by optimizing the structural parameters of the target Hamiltonian. In this work, a redefined positive phase distinct from the conventional form is employed. The positive phase is defined as an energy minimization evolution initialized from the training sample state, which drives the system toward a nearby stable state under the current parameters. This serves as a refined initialization for the subsequent negative phase Gibbs sampling. It is found that, first, the model exhibits superior performance using only a single layer $(p=1)$ in the QAOA circuit, with an average probability of 0.9559 in measuring the target quantum state under noiseless conditions. Second, the model exhibits notable noise robustness. Under the typical noise level of current mainstream commercial quantum computing devices, the average probability of measuring the target quantum state reaches 0.6047; when the noise rises to a more stringent level with doubled intensity, this probability remains at 0.3859. In both scenarios, the target quantum state maintains the highest measurement probability among all detected states, with a value several times higher than that of the second-ranked state. This indicates that the model retains strong robustness even when noise meets or exceeds the upper limit of current mainstream commercial quantum computing devices. Third, under a block-by-block learning strategy with $p=1$ and only 10 measurement shots, the model consistently generates the target “qubit” grid image regardless of noise interference, demonstrating strong robustness in image generation. The fully connected QBM proposed in this work not only possesses clear theoretical quantum advantages, but also addresses the practical bottlenecks of existing commercial quantum computing hardware, and exhibits powerful generative performance. It thus provides pivotal research support for advancing the real-world applications of quantum machine learning models in the noisy intermediate-scale quantum (NISQ) era.
    
\end{abstract}

	\maketitle
	\section{Introduction}

	Boltzmann machines represent a class of undirected probabilistic generative models grounded in statistical physics principles. By explicitly modeling the joint probability distribution of data, they provide substantial value for unsupervised learning and generative modeling applications. As the ‌original and foundational form of Boltzmann machines, the fully connected architecture was formally introduced by Ackley, Hinton, and Sejnowski (1985). Its defining characteristic is the absence of topological constraints within the network, where all nodes are fully connected via bidirectional couplings\cite{ackley1985learning}. This distinctive connectivity enables direct capture of complex high-order correlations among variables and, in theory, permits approximation of arbitrary finite discrete probability distributions, rendering fully connected Boltzmann machines one of the most expressive families of energy-based generative models\cite{leroux2008representational}.

	Despite their strong theoretical merits, fully connected Boltzmann machines incur prohibitive computational costs. Exact evaluation of the partition function requires exhaustive enumeration of node state configurations, whose complexity scales exponentially with the number of nodes, making precise inference infeasible even for moderately sized networks. Meanwhile, Gibbs sampling exhibits extremely slow convergence over fully connected topologies, requiring extensive iterations to approach the stationary distribution. This process is both time-consuming and unreliable in accuracy\cite{tieleman2008training}. Constrained by these bottlenecks, fully connected Boltzmann machines have long remained confined to theoretical validation since their inception and are rarely deployable in practical applications.

	To address these limitations, researchers have developed partially connected Boltzmann machines by imposing topological constraints, among which the most influential variants are the restricted Boltzmann machine (RBM) and the deep Boltzmann machine (DBM). The RBM was introduced by Smolensky (1986); its key innovation lies in its bipartite graph structure, which retains only connections between the visible and hidden layers while forgoing intralayer connections among nodes\cite{smolensky1986information}. This simplification enables parallel implementation of Gibbs sampling, substantially reducing the computational complexity of inference and training. Hinton (2002) proposed the contrastive divergence algorithm, which further improved the training efficiency of RBMs\cite{hinton2002training}. Hinton, Osindero, and Teh (2006) constructed the deep belief network (DBN) by stacking RBMs layer by layer\cite{hinton2006fast}. Le Roux and Bengio (2008) theoretically proved that RBMs also possess universal approximation capabilities given a sufficient number of hidden-layer nodes\cite{leroux2008representational}. Building on RBMs, Salakhutdinov and Hinton (2009, 2012) further developed DBMs and proposed efficient training schemes for such architectures\cite{salakhutdinov2009deep,salakhutdinov2012efficient}. Unlike DBNs, DBMs contain no directed connections and adhere more closely to the original statistical physics framework of Boltzmann machines, enabling more stable modeling of joint distributions and exhibiting stronger representational power in tasks including high-dimensional data generation and text modeling. Nevertheless, both RBMs and DBMs are unable to capture high-order interactions between variables due to their intralayer-connection-free topological constraints, resulting in inferior expressive power relative to fully connected architectures when handling high-dimensional, strongly correlated complex data\cite{montufar2014number}. Partially connected structures trade topological complexity for engineering practicability but fall short of adequately modeling intricate correlations within data\cite{deng2025interaction}.

	Fully connected Boltzmann machines are widely recognized for their solid theoretical foundations. Their unconstrained topology fully preserves correlation information among variables, enables far more accurate characterization of complex data distributions than partially connected architectures, and thus confers irreplaceable advantages in scenarios including high-dimensional multimodal data modeling and strongly correlated feature learning\cite{hinton2005what}. However, under classical computing frameworks, traversal of the exponentially sized state spaces is computationally prohibitive, presenting insurmountable barriers to partition function evaluation, Gibbs sampling, and large-scale parameter updates. Even with substantial modern advances in computational power, classical hardware remains fundamentally incapable of resolving the computational challenges of fully connected Boltzmann machines\cite{koller2009probabilistic}. Quantum computing offers a fundamentally new paradigm for overcoming this dilemma. Deep structural compatibility exists between quantum computing and Boltzmann machines; the former's properties of superposition and entanglement enable parallel traversal of exponential state spaces with polynomial resources. Furthermore, quantum computing frameworks can solve energy minimization problems at speeds vastly superior to classical algorithms, directly replacing conventional Gibbs sampling procedures. As highlighted by Wiebe et al. (2014), quantum architectures liberate Boltzmann machines from reliance on restricted topologies, revealing the theoretical feasibility and expressive advantages of fully connected structures, and laying a solid foundation for overcoming the classical computational bottlenecks of Boltzmann machines from the underlying physical mechanism\cite{wiebe2014quantum}.

	Quantum Boltzmann machines (QBMs) represent an integration of Boltzmann machines with quantum computing, wherein probability distributions are encoded by quantum states, and inference and learning processes are realized through quantum evolution and measurements. Wiebe and Wossnig (2019) developed two fully quantum generative training methods to resolve gradient computation difficulties for QBMs by adopting quantum relative entropy as the optimization objective\cite{wiebe2019generative}. Currently, two dominant paradigms for QBMs have emerged: quantum annealing and gate-based quantum circuits, both operating under a quantum-classical hybrid framework. This hybrid approach represents the mainstream direction in contemporary QBM research; its core principle is to leverage the complementary strengths of quantum and classical computing to optimize inference, accelerate learning procedures, and enhance the performance of Boltzmann machines\cite{srivastava2023generative}. Specifically, quantum annealing-based QBMs map classical weights to the coupling strengths and external field parameters of Ising models, where quantum annealing drives the initial quantum state toward the ground state to realize energy minimization and probabilistic sampling. In contrast, gate-based variational QBMs construct parameterized quantum circuits (PQCs) from universal quantum gate sets, and output classical samples through measurement after parameter optimization. For example, Zoufal, Lucchi, and Woerner (2020) proposed variational QBMs that adopt variational quantum imaginary time evolution to prepare Gibbs states on gate-based quantum hardware\cite{zoufal2020variational}.

	Research on QBMs has advanced rapidly in recent years. Coopmans and Benedetti (2024) established tight sample complexity bounds for QBM learning, providing theoretical guidance for model complexity design\cite{coopmans2024sample}. Patel et al. (2024) designed a novel QBM based on parameterized thermal states and hybrid quantum-classical architecture to address Hamiltonian ground-state energy estimation, enabling efficient gradient computation\cite{piatkowski2024quantum}. Demidik et al. (2025) proposed the semi-quantum RBM (sqRBM), a trainable quantum generative model that outperforms classical RBMs in hidden-unit efficiency, requiring just one-third the hidden units to learn the same distribution\cite{demidik2025expressive}. Kimura, Kato, and Hayashi (2025) examined the exponential decay of gradients induced by barren plateau in QBM training and proposed a quantum expectation maximization algorithm to circumvent nonconvex gradient optimization\cite{kimura2025structured}. Rule and Rrapaj (2025) employed unitary block encodings of RBMs to exactly represent thermal states of many-body quantum Hamiltonians and realized non-unitary operations via Trotterization of the imaginary-time propagator, addressing the problem of thermal state generation during QBM training\cite{rule2025exact}.

	Despite these remarkable advances, a survey of existing literature reveals that current research predominantly focuses on the quantum implementation of RBMs, leaving the potential of quantum computing in supporting fully connected architectures largely unexplored; studies on fully connected QBMs remain scarce. Meanwhile, in-depth investigations into the design and implementation of QBMs based on QAOA circuits, which have been validated in engineering practice and exhibit high reliability, remain rare in the literature. Furthermore, in the current NISQ era, quantum hardware noise substantially degrades sampling accuracy and fitting fidelity, making model robustness critically important. However, this aspect has not received adequate attention in existing research.

	To address this gap, this work seeks to overcome the inherent constraint of fixed target Hamiltonian in conventional QAOA circuits by treating its structural parameters as optimizable variables, aiming to construct a fully connected QBM based on bilevel optimization QAOA architecture. The subsequent research logic is structured as follows. First, the classical solution procedures of fully connected Boltzmann machines are elaborated to lay a foundation for the subsequent quantum bilevel optimization scheme. Second, a quantum computing architecture is proposed for the bilevel optimization of fully connected Boltzmann machines. Specifically, the energy function is mapped to a target Hamiltonian with optimizable structural parameters; inner-loop training simulates positive phase energy minimization via conventional QAOA circuit evolution, while outer-loop training simulates negative phase contrastive divergence learning by optimizing the structural parameters of the target Hamiltonian. Third, using a 4-qubit circuit as a case study, a detailed analysis is presented on the construction process and working principle of the fully connected QBM. Fourth, three groups of numerical experiments are conducted on the 4-qubit fully connected QBM to evaluate model convergence under different QAOA circuit depths, characterize its robustness against varying noise intensities, and assess its generative performance on image generation tasks. All source code, running logs, and experimental data are properly archived in accordance with the reproducibility principle.
	
	\section{Classical Solution Procedures for Fully Connected Boltzmann Machines}

	The Boltzmann machine investigated in this work is a fully connected Boltzmann machine. As an energy-based generative model, it is constructed based on the framework of statistical physics. To effectively capture the high-order correlations and dependencies among state variables, the model adopts a fully connected topology. This fully connected design endows the model with strong representation and approximation capabilities, enabling it to adapt to complex, multimodal, and nonlinear target distributions. Core components of the model, including its probability distribution and gradient update mechanism, have clear mathematical and physical correspondences, yielding significantly superior theoretical interpretability compared to many black-box models.

	Let $\mathcal{K} = \{k_1, k_2, \dots, k_N\}$ denote the set of all nodes in the fully connected Boltzmann machine, where the number of nodes is $|\mathcal{K}| = N$. The corresponding edge set is defined as:
	\[
	\mathcal{E} = \{(u, v) \mid u \in \mathcal{K},\, v \in \mathcal{K},\, u \neq v\}
	\]

	Let $\mathcal{B} = \{b_i \mid i=1,2,\dots,N\}$ denote the set of node biases, and let $\mathcal{W} = \{w_{ij} \mid i,j=1,2,\dots,N, i\neq j\}$ denote the set of edge weights. The overall parameter set is then given by $\varTheta = \mathcal{B} \cup \mathcal{W}$.
	Each node $k_i$ takes a binary value of $\pm 1$, denoted as $x_i \in \{\pm 1\}$, to facilitate subsequent quantum computation. Accordingly, the state vector of the Boltzmann machine is denoted by $\boldsymbol{s} = (x_1, x_2, \dots, x_N)$, and the energy function for an arbitrary state $\boldsymbol{s}$ is defined as:
	\begin{equation}
	\label{eq1}
	E(\boldsymbol{s};\varTheta) = \sum_{i=1}^{N} b_i x_i + \sum_{i=1}^{N} \sum_{j=i+1}^{N} w_{ij} x_i x_j
	\end{equation}
	
	The corresponding probability distribution function is given by:
	\begin{equation}
	\label{eq2}
	P(\boldsymbol{s}; \varTheta) = \frac{1}{Z} \exp\left(
	-\frac{1}{k_B T} \left(
	\sum_{i=1}^{N} b_i x_i + \sum_{i=1}^{N} \sum_{j=i+1}^{N} w_{ij} x_i x_j
	\right)
	\right)
	\end{equation}
	
	where $k_{\text{B}}$ is the Boltzmann constant and $T$ is the thermodynamic temperature. For computational convenience, set $k_{\text{B}} T = 1$, which simplifies the above expression to:
	\begin{equation}
	\label{eq3}
	P(\boldsymbol{s}; \varTheta) = \frac{1}{Z} \exp\left(
	-\left(
	\sum_{i=1}^{N} b_i x_i + \sum_{i=1}^{N} \sum_{j=i+1}^{N} w_{ij} x_i x_j
	\right)
	\right)
	\end{equation}
	
	where $Z$ denotes the partition function, which serves to ensure that the computed result satisfies the normalization condition. The explicit expression for $Z$ is given by:
	\begin{equation}
	\label{eq4}
	Z = \sum_{x_i, x_j \in \{\pm 1\}} \exp\left(
	-\left(
	\sum_{i=1}^{N} b_i x_i + \sum_{i=1}^{N} \sum_{j=i+1}^{N} w_{ij} x_i x_j
	\right)
	\right)
	\end{equation}
	
	Equation~(\ref{eq4}) shows that the partition function sums the Boltzmann weights of all possible node states, thereby normalizing the probability distribution over all configurations. For $N$ nodes, there exist $2^N$ distinct configurations. Consequently, in practical implementation, the computational complexity of the partition function $Z$ grows exponentially with the number of nodes. When the number of nodes becomes sufficiently large, existing classical hardware can no longer perform efficient computation. Nevertheless, as will be shown later in this work, the inherent properties of quantum computation can be fully exploited to perfectly circumvent this bottleneck.

	Given a target distribution $P_{\text{T}}(\boldsymbol{s})$, the core problem addressed by the fully connected Boltzmann machine is to obtain the optimal parameters through model training and optimization, so as to yield the optimal distribution $P(\boldsymbol{s}; \varTheta^*)$ that precisely approximates the target distribution $P_{\text{T}}(\boldsymbol{s})$. This is expressed as:
	\begin{equation}
	\label{eq5}
	\begin{gathered}
	\varTheta^* = \arg\min_{\varTheta} \mathcal{L}\left(P_{\mathrm{T}}(\boldsymbol{s}), P(\boldsymbol{s};			\varTheta)\right) \\[6pt]
	P(\boldsymbol{s};\varTheta^*) \rightarrow P_{\mathrm{T}}(\boldsymbol{s})
	\end{gathered}
	\end{equation}
	
	where $\mathcal{L}(P_{\text{T}}(\boldsymbol{s}), P(\boldsymbol{s}; \varTheta))$ denotes the loss function that quantifies the discrepancy between the model distribution and the target distribution. In practice, parameters are typically updated via Gibbs sampling-based training to obtain the optimal parameter set $\varTheta^*$. This process is decomposed into two computational procedures: positive phase energy minimization inference and negative phase contrastive divergence learning. In this work, the positive phase is defined as an energy minimization evolution initialized from the training sample state, which drives the system toward a nearby stable state under the current parameters. This serves as a refined initialization for the subsequent negative phase Gibbs sampling. The data-dependent expectation is estimated directly from the empirical mean of the training set, while the model-dependent expectation is approximated via Gibbs sampling in the negative phase.

	First, the fully connected Boltzmann machine is randomly initialized with parameter set $\varTheta^{(0)}$. Subsequently, a set of node states $\boldsymbol{s}^{(0)} = \bigl(x_1^{(0)}, x_2^{(0)}, \dots, x_N^{(0)}\bigr)$ is randomly sampled from the training data and fed into the initialized fully connected Boltzmann machine. Through iterative updates, the positive phase optimal node state $\boldsymbol{s}^{(0)^*} = \bigl(x_1^{(0)^*}, x_2^{(0)^*}, \dots, x_N^{(0)^*}\bigr)$ that achieves the minimum energy state of the system is obtained, where $\boldsymbol{s}^{(0)^*}$ satisfies:
	\begin{equation}
	\label{eq6}
	\boldsymbol{s}^{(0)*} = \arg\min_{\boldsymbol{s}} E\left(\boldsymbol{s}; \varTheta^{(0)}\right)
	\end{equation}
	
	Theoretically, the aforementioned process can only be implemented by enumerating all possible node states, thereby identifying the node state with the minimum energy. There are a total of $2^N$ possible node value combinations, and the corresponding computational complexity increases exponentially as $N$ increases. When $N$ exceeds a certain threshold, existing classical computation schemes will encounter a bottleneck. A quantum solution that has the potential to overcome this bottleneck will be presented later in this work.

	Building on this, taking $\boldsymbol{s}^{(0)^*}$ as the starting point, convergence is achieved after $G$ steps of Gibbs sampling, thereby obtaining the negative phase sample $\boldsymbol{s}^{(G)}$ under the steady state of the model distribution. Specifically, all node states can be updated sequentially using the sigmoid activation function:
	\begin{equation}
	\label{eq7}
	P(x_i = 1 \mid x_{-i}) = \sigma\left(
	2\left(
	b_i + \sum_{j \neq i} w_{ij} x_j
	\right)
	\right)
	\end{equation}
	
	In Equation~(\ref{eq7}), $P(x_i = 1 | x_{-i})$ denotes the conditional probability that the $i$-th node takes the value 1, given that the values of all other nodes remain unchanged. Accordingly, the probability that the $i$-th node takes the value $-1$ while the values of all other nodes remain unchanged is $1 - P(x_i = 1 | x_{-i})$. When the system reaches a steady state, the update process terminates, and the negative phase sample at this point is given by:
	\[
	\boldsymbol{s}^{(G)} = \left(x_1^{(G)}, x_2^{(G)}, \dots, x_N^{(G)}\right)
	\]
	
	Since each node is connected to $N-1$ other nodes, the computational complexity of a single sweep is $O(N^2)$. As the number of nodes $N$ increases, the computational cost per iteration expands rapidly. Meanwhile, multiple sweeps are required for the system to reach a steady state, and the corresponding total computational complexity is further amplified to $O(G \times N^2)$. As $N$ increases, the computational difficulty grows explosively. In practice, when $N > 100$, the classical sampling process can hardly converge within a reasonable time. Later in this work, quantum computing principles will be employed to optimize this process.

	For the loss function, Kullback-Leibler (KL) divergence is commonly adopted, whose expression is given by:
	\begin{equation}
	\label{eq8}
	\mathcal{L}\left(P_{\mathrm{T}}(\boldsymbol{s}), P(\boldsymbol{s};\varTheta)\right)
	= D_{\mathrm{KL}}\left(P_{\mathrm{T}}(\boldsymbol{s}) \parallel P(\boldsymbol{s};\varTheta)\right)
	= \mathbb{E}_{P_{\mathrm{T}}(\boldsymbol{s})} \left[
	\log \frac{P_{\mathrm{T}}(\boldsymbol{s})}{P(\boldsymbol{s};\varTheta)}
	\right]
	\end{equation}
	
	Based on this, the gradients of the parameters $b_i$ and $w_{ij}$ can be further derived :
	\begin{equation}
	\label{eq9}
	\begin{gathered}
	\frac{\partial \mathcal{L}}{\partial b_i}
	= -\left(
	\mathbb{E}_{P_{\mathrm{T}}(\boldsymbol{s})} \left[ x_i \right]
	- \mathbb{E}_{P(\boldsymbol{s};\varTheta)} \left[ x_i \right]
	\right) \\[10pt]
	\frac{\partial \mathcal{L}}{\partial w_{ij}}
	= -\left(
	\mathbb{E}_{P_{\mathrm{T}}(\boldsymbol{s})} \left[ x_i x_j \right]
	- \mathbb{E}_{P(\boldsymbol{s};\varTheta)} \left[ x_i x_j \right]
	\right)
	\end{gathered}
	\end{equation}
	
	where the data-dependent expectation $\mathbb{E}_{P_\text{T}(s)}[\cdot]$ is estimated via the empirical mean over the training dataset, while the model-dependent expectation $\mathbb{E}_{P(s; \varTheta)}[\cdot]$ is approximated by Gibbs sampling. Based on the gradients, by setting the learning rate $\eta_1$, the parameters of the fully connected Boltzmann machine can be updated using gradient descent:
	\begin{equation}
	\label{eq10}
	\begin{gathered}
	b_i^{(t+1)} \leftarrow b_i^{(t)} + \eta_1 \left(
	\mathbb{E}_{P_{\mathrm{T}}(\boldsymbol{s})} \left[ x_i \right]
	- \mathbb{E}_{P(\boldsymbol{s};\varTheta)} \left[ x_i \right]
	\right) \\[10pt]
	w_{ij}^{(t+1)} \leftarrow w_{ij}^{(t)} + \eta_1 \left(
	\mathbb{E}_{P_{\mathrm{T}}(\boldsymbol{s})} \left[ x_i x_j \right]
	- \mathbb{E}_{P(\boldsymbol{s};\varTheta)} \left[ x_i x_j \right]
	\right)
	\end{gathered}
	\end{equation}
	
	Through multiple iterations, the parameter values gradually approach the optimal parameter set $\varTheta^*$, and the distribution generated by the model also progressively approximates the target distribution. Despite its numerous advantages, the training and computational implementation of the fully connected Boltzmann machine still suffer from significant bottlenecks, which greatly limit its practical performance in real-world scenarios. For example, to compute various expectations under the model distribution, approximation via Gibbs sampling is required. The sampling procedure is not only time-consuming and computationally expensive but also highly prone to problems including mode collapse. In addition, the intractable partition function prohibits the direct analytical derivation of gradients, rendering sampling-based approximation the standard approach. As the state dimension rises, the computational overhead of sampling and gradient updates increases exponentially. This work further explores quantum computing principles to mitigate these bottlenecks.

	\section{Bilevel Optimization Quantum Computing Process for Fully Connected Boltzmann Machines}

	To address the bottlenecks in classical computation, this work attempts to optimize the solving process of fully connected Boltzmann machines using quantum computing principles. As revealed by the analysis and derivation of the classical solving procedure in the preceding sections, the bottlenecks are concentrated in two core steps: positive phase energy minimization inference and negative phase contrastive divergence learning. In view of this, this work breaks through the conventional constraint that the target Hamiltonian of the QAOA circuit is fixed, treats its structural parameters as optimizable variables, and reformulates the above two steps within the framework of quantum computing. Specifically, first, the QAOA circuit is used to simulate the positive phase energy minimization process, advancing the inefficient point-by-point sampling updates to synchronous updates over all qubit nodes, and the ground-state configuration of the system is obtained via the parameter feedback loop of the QAOA circuit. Second, reverse evolution is performed on the QAOA circuit to simulate the negative phase contrastive divergence learning process, replacing the inefficient sampling and parameter updates with direct measurement and update of the quantum state, thereby yielding the optimal parameter set that accurately fits the target distribution.

	Theoretically, the fully connected Boltzmann machine based on the QAOA architecture proposed in this work can fully exploit the parallelism and entanglement inherent to quantum computing. This not only significantly accelerates node updates and sampling, but also enhances the model's expressive power and its ability to explore high-dimensional feature spaces. Furthermore, quantum measurement results naturally satisfy normalized probabilistic properties, eliminating the need for intractable partition function computations and thereby drastically simplifying the overall workflow. Notably, the QAOA-based model is executable entirely on current NISQ hardware, demonstrating strong practical engineering value.

	\subsection{Mapping Energy Function to Target Hamiltonian $H_1$}

	First, the energy function of the fully connected Boltzmann machine is mapped to the corresponding target Hamiltonian $H_1$, and the initial biases and connection weights are set.

	Following the underlying mechanism of the QAOA circuit, the energy function of the fully connected Boltzmann machine is transformed into the corresponding target Hamiltonian $H_1$. The specific transformation rules are as follows: (1) The single node $x_i$ is converted into a composite operator (denoted as $\sigma_{\text{z}}^i$) that only applies the Pauli-Z gate to the $i$-th qubit and acts trivially on other qubits; (2) The node product term $x_ix_j$ is transformed into a composite operator $\sigma_{\text{z}}^i\sigma_{\text{z}}^j$ that applies the Pauli-Z gate to both the $i$-th and $j$-th qubits and acts trivially on other qubits; (3) The parameter set $\varTheta$ consisting of biases $b_i$ and connection weights $w_{ij}$ is directly incorporated into the target Hamiltonian. According to the above rules, the energy function $E(\boldsymbol{s}; \varTheta)$ can be converted into the corresponding target Hamiltonian $H_1$:
	\[
	E(\boldsymbol{s};\varTheta) \rightarrow H_1 = \sum_{i=1}^{N} b_i \sigma_\text{z}^i + \sum_{i=1}^{N} \sum_{j=i+1}^{N} w_{ij} \sigma_\text{z}^i \sigma_\text{z}^j
	\]

	From the above equation, the $N$ nodes of the fully connected Boltzmann machine are mapped to $N$ qubits, and the node variables in the energy polynomial are converted into composite operators that apply the Pauli-Z gate only to specific qubits, i.e., $\sigma^i_{\text{z}}$ or $\sigma^i_{\text{z}}\sigma^j_{\text{z}}$. The Pauli-Z gate satisfies the following properties:
	\[
	\sigma_\text{z} |0\rangle = |0\rangle, \quad \sigma_\text{z} |1\rangle = -|1\rangle
	\]

	In this way, the two states of the nodes in the classical fully connected Boltzmann machine, which take values of $\pm 1$, are mapped to the corresponding qubit states $|0\rangle$ and $|1\rangle$, thus laying the foundation for subsequent processing via quantum computing.

	\subsection{Inner-Loop Training: Simulating Positive Phase Energy Minimization Based on QAOA Computation}

	Subsequently, the positive phase energy minimization process of the fully connected Boltzmann machine is simulated based on the mechanism of the QAOA circuit.

	Let $H(t)$ be a time-dependent Hamiltonian. By solving the Schrödinger equation, the corresponding quantum state evolution equation is obtained as follows:
	\[
	|\varphi(t)\rangle = \exp\left(-\frac{\text{i} H(t)}{\hbar}\right) |\varphi(0)\rangle
	\]

	Setting the reduced Planck constant $\hbar=1$, the quantum state evolution equation is simplified to:
	\begin{equation}
	\label{eq11}
	|\varphi(t)\rangle = \exp\left(-\text{i} H(t)\right) |\varphi(0)\rangle
	\end{equation}

	According to the quantum adiabatic algorithm\cite{farhi2000quantum}, the Hamiltonian $H(t)$ is further decomposed into:
	\begin{equation}
	\label{eq12}
	H(t) = \left(1 - \frac{t}{T}\right) H_0 + \frac{t}{T} H_1
	\end{equation}

	where $T$ is the total evolution time and $H_0$ is the initial Hamiltonian. Based on the QAOA solution principle, the aforementioned evolution process can be represented through a discretized approximation approach\cite{farhi2014quantum}:
	\begin{equation}
	\label{eq13}
	\exp\left(\text{i}\beta_p H_0\right)\exp\left(\text{i}\gamma_p H_1\right)\dots
	\exp\left(\text{i}\beta_2 H_0\right)\exp\left(\text{i}\gamma_2 H_1\right)
	\exp\left(\text{i}\beta_1 H_0\right)\exp\left(\text{i}\gamma_1 H_1\right)|\varphi(0)\rangle
	\end{equation}

	In Equation~(\ref{eq13}), $p$ denotes the number of discrete sub-processes into which the continuous evolution is decomposed, i.e., the depth of the QAOA circuit. The corresponding vectors $\boldsymbol{\beta} = (\beta_1, \beta_2, \dots, \beta_p)$ and $\boldsymbol{\gamma} = (\gamma_1, \gamma_2, \dots, \gamma_p)$ represent two sets of variational parameters. If the entire circuit contains $N$ qubits, the initial quantum state is given by:
	\begin{equation}
	\label{eq14}
	|\varphi(0)\rangle = \mathrm{H}^{\otimes N} |0\rangle^{\otimes N} = |+\rangle^{\otimes N} = \frac{1}			{\sqrt{2^N}} \sum_{i=0}^{2^N-1} |i\rangle
	\end{equation}

	where $\text{H}$ denotes the Hadamard gate. The initial Hamiltonian $H_0$ is defined as:
	\[
	H_0 = \sum_{i=1}^{N} \sigma_\text{x}^i
	\]

	where $\sigma_{\text{x}}^i$ denotes the application of the Pauli-X gate on the $i$-th qubit, while the identity operator I is applied to the remaining qubits. After constructing the QAOA circuit according to the aforementioned solution process, the initial values of the vectors $\boldsymbol{\beta}$ and $\boldsymbol{\gamma}$ are first determined randomly. These values are then fed into the circuit to implement the approximate quantum evolution process and perform measurements, thereby yielding the expected energy corresponding to the target Hamiltonian $H_1$:
	\begin{equation}
	\label{eq15}
	\langle H_1\rangle
	= \langle \boldsymbol{\gamma}, \boldsymbol{\beta} | H_1 | \boldsymbol{\gamma}, \boldsymbol{\beta} 				\rangle
	= \sum_{i=0}^{2^N-1} |\alpha_i|^2 \langle i | H_1 | i \rangle
	\approx \sum_{i=0}^{2^N-1} \frac{m_i}{m} \langle i | H_1 | i \rangle
	\end{equation}

	where $m$ denotes the total number of measurements, and $m_i$ denotes the number of times the quantum state $|i\rangle$ is measured. Based on the expected energy computed via Equation~(\ref{eq15}), the corresponding gradient can be obtained through classical or quantum computation. The values of the vectors $\boldsymbol{\beta}$ and $\boldsymbol{\gamma}$ are then updated accordingly until the expected energy converges. In general, the parameter-shift rule\cite{mitarai2018quantum}, which is based on quantum computing principles, provides substantial advantages for parameter updates in quantum circuits. For any parameter $\beta_k$ in the vector $\boldsymbol{\beta}$, the gradient derived from the parameter-shift rule is given by:
	\begin{equation}
	\label{eq16}
	\begin{aligned}
	\frac{\partial \langle H_1 \rangle}{\partial \beta_k}
	&= \frac{1}{2} \left(
	\langle H_1 \rangle\left(\beta_k + \frac{\pi}{2}\right)
	- \langle H_1 \rangle\left(\beta_k - \frac{\pi}{2}\right)
	\right) \\
	&\approx \frac{1}{2} \left(
	\sum_{i=0}^{2^N-1} \frac{m_i^{\beta^+}}{m} \langle i | H_1 | i \rangle
	- \sum_{i=0}^{2^N-1} \frac{m_i^{\beta^-}}{m} \langle i | H_1 | i \rangle
	\right)
	\end{aligned}
	\end{equation}

	In Equation~(\ref{eq16}), $\langle H_1\rangle(\beta_k+\pi/2)$ and $\langle H_1\rangle(\beta_k-\pi/2)$ denote the expected energies of the target Hamiltonian $H_1$ with the parameter $\beta_k$ perturbed by $\pm\pi/2$. The terms $m_i^{\beta^+}$ and $m_i^{\beta^-}$ denote the measurement counts for each basis bitstring under the $\pm\pi/2$ perturbations of $\beta_k$. Based on the obtained gradient, the parameter $\beta_k$ is updated via a classical optimization algorithm:
	\[
	\beta_k^{(t+1)} \leftarrow \beta_k^{(t)} - \eta_2 \frac{\partial \langle H_1 \rangle}{\partial 					\beta_k}
	\]

	where $\eta_2$ denotes the learning rate. The gradient for any parameter $\gamma_k$ in the vector $\boldsymbol{\gamma}$ is given by:
	\begin{equation}
	\label{eq17}
	\begin{aligned}
	\frac{\partial \langle H_1 \rangle}{\partial \gamma_k}
	&= \frac{1}{2} \left(
	\langle H_1 \rangle\left(\gamma_k + \frac{\pi}{2}\right)
	- \langle H_1 \rangle\left(\gamma_k - \frac{\pi}{2}\right)
	\right) \\
	&\approx \frac{1}{2} \left(
	\sum_{i=0}^{2^N-1} \frac{m_i^{\gamma^+}}{m} \langle i | H_1 | i \rangle
	- \sum_{i=0}^{2^N-1} \frac{m_i^{\gamma^-}}{m} \langle i | H_1 | i \rangle
	\right)
	\end{aligned}
	\end{equation}

	In Equation~(\ref{eq17}), $\langle H_1\rangle(\gamma_k+\pi/2)$ and $\langle H_1\rangle(\gamma_k-\pi/2)$ denote the expected energies of the target Hamiltonian $H_1$ with the parameter $\gamma_k$ perturbed by $\pm\pi/2$. The terms $m_i^{\gamma^+}$ and $m_i^{\gamma^-}$ denote the measurement counts for each basis bitstring under the $\pm\pi/2$ perturbations of $\gamma_k$. With this gradient, the parameter $\gamma_k$ is updated via a classical algorithm:
	\[
	\gamma_k^{(t+1)} \leftarrow \gamma_k^{(t)} - \eta_2 \frac{\partial \langle H_1 \rangle}{\partial 				\gamma_k}
	\]

	The above parameter updating process is repeated until the expected energy converges. At this point, the optimized parameter vectors $\boldsymbol{\beta}^*$ and $\boldsymbol{\gamma}^*$ are incorporated into the QAOA circuit for execution, yielding the approximate optimal quantum state corresponding to the target Hamiltonian $H_1$, that is, the quantum state that minimizes the energy of $H_1$ under fixed biases and connection weights.

	The above scheme leverages the QAOA circuit to simulate the positive phase energy minimization process of the fully connected Boltzmann machine. It overcomes the computational complexity bottleneck in the sampling and updating procedures of classical algorithms, and fully exploits quantum parallelism to globally compute the energy states of the entire system in a synchronous manner. Gradients are obtained via the parameter-shift rule, and the parameter vectors $\boldsymbol{\beta}$ and $\boldsymbol{\gamma}$ are updated to ultimately approximate the optimal parameter configuration that minimizes the overall system energy. Admittedly, this scheme serves only as an intermediate step for solving fully connected Boltzmann machines, with the goal of achieving the minimum system energy under the given parameter set $\varTheta$, so as to approximately simulate the positive phase energy minimization procedure. Subsequently, the parameter set $\varTheta$ of the target Hamiltonian $H_1$ requires further optimization according to the target distribution to simulate the negative phase contrastive divergence learning process. This framework differs substantially from conventional QAOA implementations, where the quantum state corresponding to the lowest energy is adopted as the final approximate solution while the target Hamiltonian parameters remain fixed.

	\subsection{Outer-Loop Training: Simulating Negative Phase Contrastive Divergence Learning via Optimizing Target Hamiltonian Parameters}

	On this basis, the negative phase contrastive divergence learning process is further simulated via the QAOA circuit. It should be noted that, considering the inherent training characteristics of a fully connected Boltzmann machine, this work attempts to break the conventional limitation of fixed target Hamiltonians in standard QAOA circuits by treating structural parameters as optimizable variables. First, the optimal parameter vectors $\boldsymbol{\beta}^*$ and $\boldsymbol{\gamma}^*$ obtained via inner-loop training are loaded into the QAOA circuit to prepare the intermediate quantum state $|\psi\rangle = |\boldsymbol{\beta}^*, \boldsymbol{\gamma}^*\rangle$, which corresponds to the node assignments that approximately minimize the system energy under fixed model parameters. Next, to reduce the discrepancy between the model distribution and the target distribution, the corresponding loss Hamiltonian is constructed. To improve computational efficiency and facilitate quantum processing, this work employs the mean squared error (MSE) to quantify the deviation between the model and target distributions. The corresponding Hamiltonian $H_2$ is defined as:
	\begin{equation}
	\label{eq18}
	H_2 = \frac{1}{2^N} \sum_{i=0}^{2^N-1} \left( M_i \rho_\psi M_i - 2 P_{\mathrm{T}}(i) M_i + P_{\mathrm{T}}(i)^2 \text{I} \right)
	\end{equation}

	where $M_i$ denotes the measurement matrix of the $i$-th basis vector, and $\rho_\psi$ represents the density matrix of the intermediate state $|\psi\rangle$:
	\[
	\begin{aligned}
	M_i = |i\rangle\langle i|, \quad \rho_\psi = |\psi\rangle\langle\psi|
	\end{aligned}
	\]

	where $P_\text{T}(i)$ denotes the basis-state probability distribution transformed from the target distribution. Specifically, each element $x_i$ in the state vector $ \boldsymbol{s}$ is mapped onto the quantum basis states according to the rule below:
	\[
	\begin{aligned}
	x_i = -1 \rightarrow |1\rangle, \quad x_i = +1 \rightarrow |0\rangle
	\end{aligned}
	\]

	The intermediate quantum state $|\psi\rangle = |\boldsymbol{\beta}^*, \boldsymbol{\gamma}^*\rangle$ is measured with the Hamiltonian $H_2$ defined in Equation~(\ref{eq18}):
	\clearpage

	\[
	\begin{aligned}
\langle H_2\rangle
&= \langle \psi | H_2 | \psi \rangle \\
&= \langle \psi | \frac{1}{2^N} \sum_{i=0}^{2^N-1} \left( M_i \rho_\psi M_i - 2 P_{\mathrm{T}}(i) M_i + P_{\mathrm{T}}(i)^2 \text{I} \right) | \psi \rangle \\
&= \frac{1}{2^N} \sum_{i=0}^{2^N-1} \langle \psi | \left( |i\rangle\langle i|\psi\rangle\langle\psi|i\rangle\langle i| - 2 P_{\mathrm{T}}(i) |i\rangle\langle i| + P_{\mathrm{T}}(i)^2 \text{I} \right) | \psi \rangle \\
&= \frac{1}{2^N} \sum_{i=0}^{2^N-1} \left( \langle\psi|i\rangle\langle i|\psi\rangle\langle\psi|i\rangle\langle i|\psi\rangle - 2 P_{\mathrm{T}}(i) \langle\psi|i\rangle\langle i|\psi\rangle + P_{\mathrm{T}}(i)^2 \langle\psi|\text{I}|\psi\rangle \right) \\
&= \frac{1}{2^N} \sum_{i=0}^{2^N-1} \left( \left( \langle\psi|M_i|\psi\rangle \right)^2 - 2 P_{\mathrm{T}}(i) \langle\psi|M_i|\psi\rangle + P_{\mathrm{T}}(i)^2 \langle\psi|\psi\rangle \right) \\
&= \frac{1}{2^N} \sum_{i=0}^{2^N-1} \left( P(i)^2 - 2 P_{\mathrm{T}}(i) P(i) + P_{\mathrm{T}}(i)^2 \right) \\
&= \frac{1}{2^N} \sum_{i=0}^{2^N-1} \left( P(i) - P_{\mathrm{T}}(i) \right)^2
	\end{aligned}
	\]

	where $P(i)$ denotes the probability of measuring state $i$. It is clear that measuring the intermediate quantum state with the Hamiltonian $H_2$ yields the MSE between the model distribution and the target distribution. In practice, to further simplify the computation, the MSE can be approximated via a computational procedure that directly measures the probability distribution over all basis vectors:
	\begin{equation}
	\label{eq19}
\langle H_2 \rangle = \mathrm{MSE}
= \frac{1}{2^N} \sum_{i=0}^{2^N-1} \left( P(i) - P_{\mathrm{T}}(i) \right)^2
\approx \frac{1}{2^N} \sum_{i=0}^{2^N-1} \left( \frac{m_{\psi i}}{m} - P_{\mathrm{T}}(i) \right)^2
	\end{equation}

	In Equation~(\ref{eq19}), $m$ denotes the total number of measurements, and $m_{\psi i}$ denotes the number of measurements for each basis bitstring from the intermediate state $|\psi\rangle$. The gradient of the loss function can then be derived via the parameter-shift rule, and the biases $b_i$ and connection weights $w_{ij}$ in  $H_2$ are updated accordingly:
	\begin{equation}
	\label{eq20}
	\begin{aligned}
\frac{\partial \langle H_2 \rangle}{\partial b_i}
&= \frac{1}{2} \left(
\langle H_2 \rangle\left(b_i + \frac{\pi}{2}\right)
- \langle H_2 \rangle\left(b_i - \frac{\pi}{2}\right)
\right) \\
&\approx \frac{1}{2} \left(
\sum_{i=0}^{2^N-1} \frac{m_{\psi i}^{b^+}}{m} \langle i | H_2 | i \rangle
- \sum_{i=0}^{2^N-1} \frac{m_{\psi i}^{b^-}}{m} \langle i | H_2 | i \rangle
\right)
	\end{aligned}
	\end{equation}

	\[
	b_i^{(t+1)} \leftarrow b_i^{(t)} - \eta_3 \frac{\partial \langle H_2 \rangle}{\partial b_i}
	\]

	\begin{equation}
	\label{eq21}
	\begin{aligned}
\frac{\partial \langle H_2 \rangle}{\partial w_{ij}}
&= \frac{1}{2} \left(
\langle H_2 \rangle\left(w_{ij} + \frac{\pi}{2}\right)
- \langle H_2 \rangle\left(w_{ij} - \frac{\pi}{2}\right)
\right) \\
&\approx \frac{1}{2} \left(
\sum_{i=0}^{2^N-1} \frac{m_{\psi i}^{w^+}}{m} \langle i | H_2 | i \rangle
- \sum_{i=0}^{2^N-1} \frac{m_{\psi i}^{w^-}}{m} \langle i | H_2 | i \rangle
\right)
	\end{aligned}
	\end{equation}

	\[
	w_{ij}^{(t+1)} \leftarrow w_{ij}^{(t)} - \eta_3 \frac{\partial \langle H_2 \rangle}{\partial w_{ij}}
	\]

	where $\eta_3$ is the learning rate, and $\langle H_2 \rangle(b_i \pm \pi/2)$ and $\langle H_2 \rangle(w_{ij} \pm \pi/2)$ denote the expected energies of the Hamiltonian $H_2$ with the parameters $b_i$ and $w_{ij}$ perturbed by $\pm\pi/2$, respectively. $m^{b\pm}_i$ and $m^{w\pm}_{ij}$ denote the number of measurements for each basis bitstring under the corresponding perturbation.

	The above process simulates the negative phase contrastive divergence learning of the fully connected Boltzmann machine via the QAOA circuit, derives the loss gradient using the parameter-shift rule, and updates the biases and connection weights accordingly, ultimately yielding the optimal parameter set $\varTheta^*$ for $H_1$ that minimizes the discrepancy between the model and target distributions. This framework offers two key advantages. First, by exploiting the inherent parallelism of quantum computing to globally and simultaneously compute the probabilities of all system states, it avoids the computationally costly, time-consuming, and collapse-prone Gibbs sampling in classical training. Second, it takes advantage of the intrinsic probabilistic nature of quantum measurement to directly acquire all-state probabilities, thereby elegantly circumventing the intractable partition function computation in conventional Boltzmann machine training.

	\subsection{Updating Parameters via Bilevel Training until $\langle H_2 \rangle$ Converges}

	The parameters of the fully connected quantum Boltzmann machine are updated via inner- and outer-loop training until the convergence condition for $\langle H_2 \rangle$ is met. Here, inner-loop training refers to the complete process of solving for the optimal $\boldsymbol{\beta}$ and $\boldsymbol{\gamma}$ through conventional QAOA circuit computation. With $\boldsymbol{\beta}$ and $\boldsymbol{\gamma}$ fixed, outer-loop training derives the loss gradient and updates the parameter set $\varTheta$ of $H_1$ by measuring the discrepancy between the model distribution and the target distribution. Once $\varTheta$ is updated, a new target Hamiltonian $H_1$ is constructed and embedded into the QAOA circuit for re-execution, yielding new optimal $\boldsymbol{\beta}$ and $\boldsymbol{\gamma}$, which further guides the next update of $\varTheta$. This inner- and outer-loop iterative process proceeds until the MSE loss $\langle H_2 \rangle$ converges, ultimately yielding the optimal parameter configurations $\varTheta^*$, $\boldsymbol{\beta}^*$, and $\boldsymbol{\gamma}^*$. Finally, these optimized parameters are incorporated into the fully connected QBM based on the QAOA circuit. The circuit is then executed and measured to generate a model distribution that precisely approximates the target distribution.

	It should be noted that there are two typical technical schemes for the practical implementation of the inner- and outer-loop training process. The first is the sequential scheme: in each global iteration, the outer-loop parameters are fixed initially, and the inner-loop training is then executed multiple times until convergence; subsequently, the inner-loop parameters are fixed, and the outer-loop training is executed multiple times until convergence. The second is the nested scheme: in each global iteration, the inner-loop training is first executed multiple times to obtain the current optimal inner-loop parameters, followed by a single outer-loop training step to update the outer-loop parameters, before proceeding to the next iteration. Computational experiments demonstrate that the sequential scheme significantly outperforms the nested scheme in terms of both overall performance and resource consumption. For this reason, this work adopts the sequential scheme to integrate the inner- and outer-loop training processes. In addition, in practice, a maximum number of iterations is set as a fallback mechanism to prevent infinite loops caused by nonconvergence. Training is forcibly terminated once the maximum iterations is reached, even if the loss value has not yet converged. The complete algorithm framework is presented below:
	\clearpage

	\begin{figure}[htbp]
  	\centering
 	\vspace{-80pt}
  	\hspace*{-0.14\textwidth}
  	\includegraphics[width=1.255\textwidth]{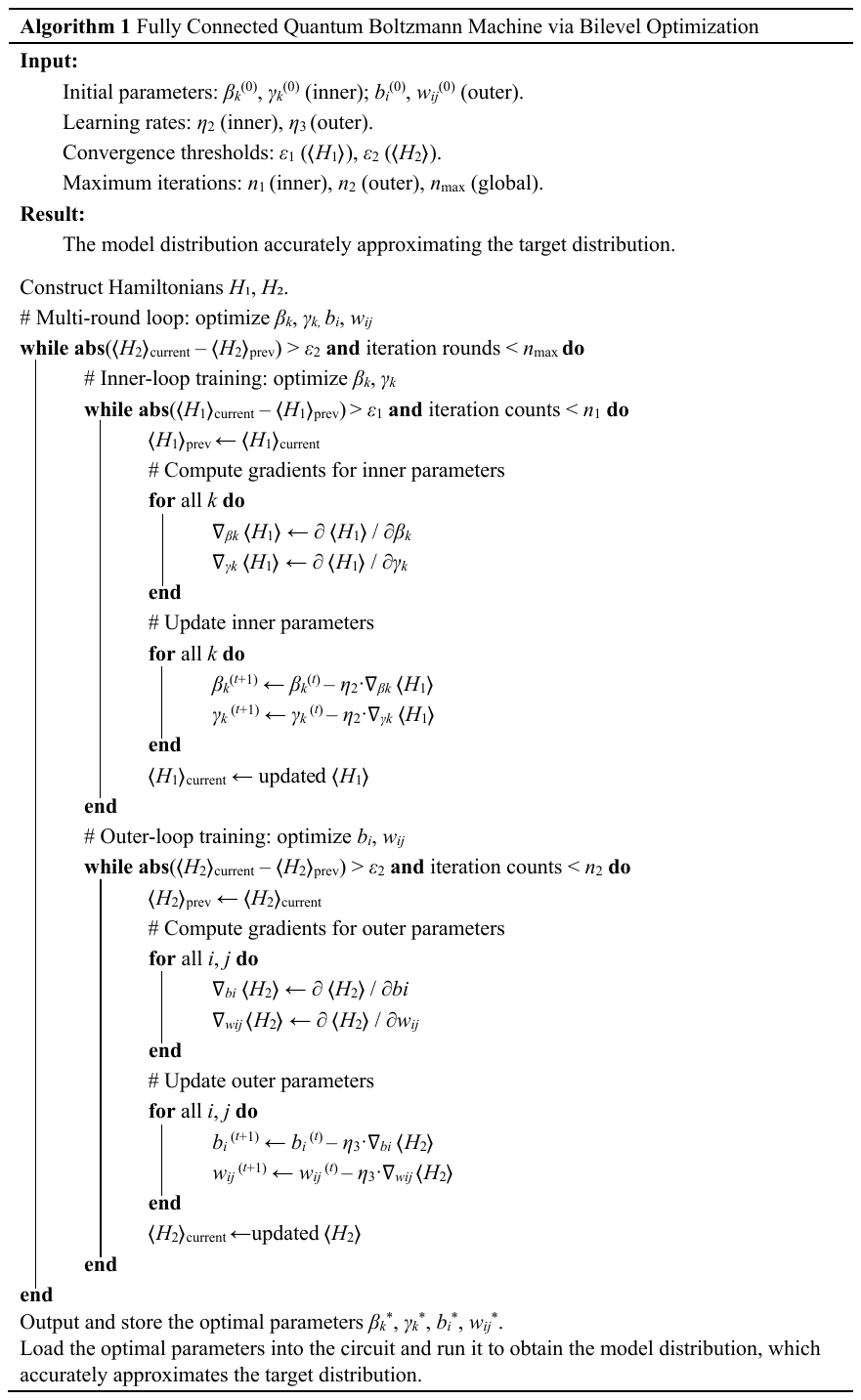}
 	\label{alg}
	\end{figure}
	\clearpage

	\section{4-Qubit Fully Connected Quantum Boltzmann Machine}

	\subsection{Construction of Quantum Computing Circuits}

	In the following, a 4-qubit quantum circuit is taken as an example to elaborate on the construction process and computational principle of the fully connected QBM. Based on the derivation in the theoretical model section, for $N=4$, the corresponding initial quantum state is prepared as follows:
	\begin{equation}
	\label{eq22}
	|\varphi(0)\rangle = H^{\otimes 4} |0\rangle^{\otimes 4} = |+\rangle^{\otimes 4} = \frac{1}{\sqrt{2^4}} \sum_{i=0}^{2^4-1} |i\rangle
	\end{equation}

	The initial Hamiltonian $H_0$ is given by:
	\begin{equation}
	\label{eq23}
	H_0 = \sigma_{\text{x}}^1 + \sigma_{\text{x}}^2 + \sigma_{\text{x}}^3 + \sigma_{\text{x}}^4
	\end{equation}

	The target Hamiltonian $H_1$ corresponding to the energy function is given by:
	\begin{equation}
	\label{eq24}
	\begin{aligned}
H_1 &= b_1 \sigma_{\text{z}}^1 + b_2 \sigma_{\text{z}}^2 + b_3 \sigma_{\text{z}}^3 + b_4 \sigma_{\text{z}}^4 + w_{12} \sigma_{\text{z}}^1 \sigma_{\text{z}}^2 + w_{13} \sigma_{\text{z}}^1 \sigma_{\text{z}}^3 \\
&\quad + w_{14} \sigma_{\text{z}}^1 \sigma_{\text{z}}^4 + w_{23} \sigma_{\text{z}}^2 \sigma_{\text{z}}^3 + w_{24} \sigma_{\text{z}}^2 \sigma_{\text{z}}^4 + w_{34} \sigma_{\text{z}}^3 \sigma_{\text{z}}^4
	\end{aligned}
	\end{equation}

	The quantum state evolution equation derived from the QAOA circuit is given by:
	\begin{equation}
	\label{eq25}
\exp(\text{i}\beta_p H_0) \exp(\text{i}\gamma_p H_1) \cdots \exp(\text{i}\beta_2 H_0) \exp(\text{i}\gamma_2 H_1) \exp(\text{i}\beta_1 H_0) \exp(\text{i}\gamma_1 H_1) |\varphi(0)\rangle
	\end{equation}

	where $p$ denotes the number of QAOA circuit layers. Accordingly, the basic structure of an arbitrary $k$-th layer can be constructed as follows:
	\begin{figure}[htbp]
  	\centering
  	\includegraphics[width=\textwidth]{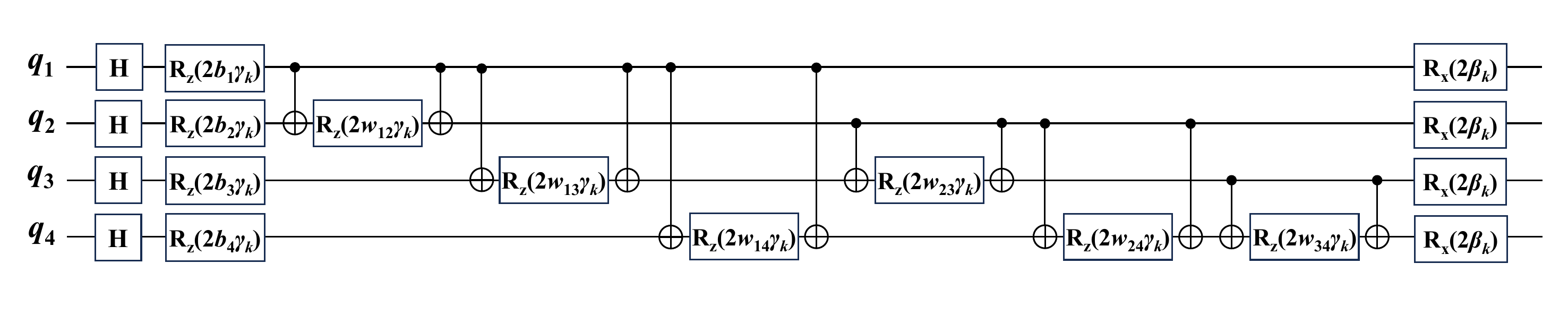}
  	\caption{Structure of the single-layer QAOA circuit for the 4-qubit fully connected quantum Boltzmann machine}
  	\label{fig1}
	\end{figure}

	It should be noted that all four qubits in Figure \ref{fig1} are initialized to the $|0\rangle$ state. On this basis, the architecture of the 4-qubit fully connected QBM is further established.
	\clearpage	
	\vspace*{0.5cm}
	\begin{figure}[htbp]
	\centering
	\includegraphics[width=\linewidth]{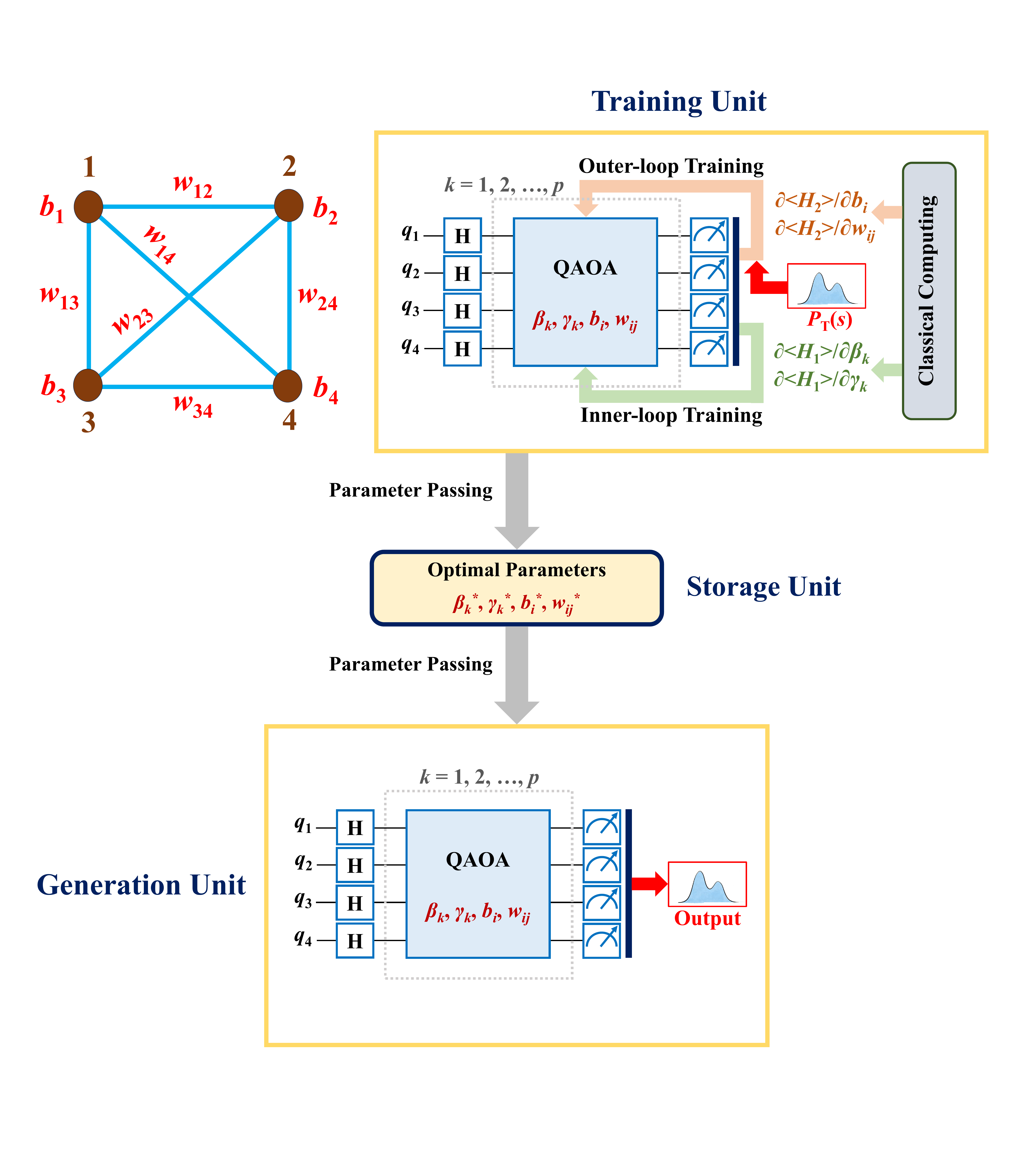}
	\caption{Architecture of the 4-qubit fully connected quantum Boltzmann machine}
	\label{fig2}
	\end{figure}
	\clearpage	

	As shown in Figure~\ref{fig2}, the fully connected QBM consists of three parts: a training unit, a generation unit, and a storage unit. The training unit updates and optimizes all parameters according to the target distribution. Based on the optimal parameters obtained from training, the generation unit outputs a model distribution that precisely approximates the target distribution. The storage unit stores and transfers the optimal parameters.

	According to the mathematical derivation, the training unit essentially implements an inner- and outer-loop training process based on a quantum-classical hybrid architecture. The inner-loop training utilizes the QAOA circuit to compute $\langle H_1 \rangle$ and its gradient, and updates the parameters $\beta_k$ and $\gamma_k$ via a classical optimization algorithm until $\langle H_1 \rangle$ is minimized. Similarly, the outer-loop training computes $\langle H_2 \rangle$ and its gradient through the QAOA circuit, and updates the parameters $b_i$ and $w_{ij}$ using the same algorithm until $\langle H_2 \rangle$ is minimized. Theoretically, by fully exploiting the parallelism and entanglement properties of quantum computing, this unit globally calculates the probabilities of all possible states simultaneously, thereby overcoming the technical bottleneck of conventional Gibbs sampling. Meanwhile, it leverages the inherent probabilistic nature of quantum measurement to directly obtain the probability distribution of each state, elegantly avoiding the intractable partition function computation problem. As illustrated in Figure~\ref{fig2}, the peripheral classical computing module handles the parameter updates for the training unit.

	The generation unit adopts the same QAOA circuit structure as the training unit. After receiving the optimal parameters from the training unit, the generation unit assigns these parameters to the corresponding variables in its QAOA circuit. By executing the quantum circuit and implementing measurements, it yields a model distribution that precisely approximates the target distribution. The storage unit centrally stores the optimal parameters generated by the training unit, enabling convenient access, retrieval, and export at any stage. It thus acts as an information bridge, facilitating seamless parameter transfer between the training unit and the generation unit.

	\subsection{Setting the Target Probability Distribution}

	For a 4-qubit circuit, there are a total of $2^4 = 16$ possible measurement outcomes, corresponding to 16 mutually orthogonal basis vectors. To match these 16 results of the 4-qubit measurements and thus facilitate the computation of the MSE loss, this work expands all possible values of the target distribution into a 16-dimensional column vector via one-hot encoding, as shown in Table~\ref{tab1}.
	\begin{table}[htbp]
  	\centering
  	\caption{One-hot encoding of target values}
  	\label{tab1}
  	\begin{tabular}{c@{\hspace{48pt}}c} 
   \noalign{\hrule height 0.8pt}
    Target values & One-hot vectors\\
   \hline
    0000 & $[0000000000000001]^\text{T}$ \\
    0001 & $[0000000000000010]^\text{T}$ \\
    0010 & $[0000000000000100]^\text{T}$ \\
    0011 & $[0000000000001000]^\text{T}$ \\
    0100 & $[0000000000010000]^\text{T}$ \\
    0101 & $[0000000000100000]^\text{T}$ \\
    0110 & $[0000000001000000]^\text{T}$ \\
    0111 & $[0000000010000000]^\text{T}$ \\
    1000 & $[0000000100000000]^\text{T}$ \\
    1001 & $[0000001000000000]^\text{T}$ \\
    1010 & $[0000010000000000]^\text{T}$ \\
    1011 & $[0000100000000000]^\text{T}$ \\
    1100 & $[0001000000000000]^\text{T}$ \\
    1101 & $[0010000000000000]^\text{T}$ \\
    1110 & $[0100000000000000]^\text{T}$ \\
    1111 & $[1000000000000000]^\text{T}$ \\
   \noalign{\hrule height 0.8pt}
	\end{tabular}
	\end{table}

	In practical computational experiments, the measurement outcomes obtained after executing the quantum circuit are naturally presented as 16-dimensional column vectors, whose dimensionality exactly matches that of the one-hot vectors in Table~\ref {tab1}. Furthermore, to simplify subsequent computation, the target distribution in this work is set as a one-point distribution. That is, the probability of sampling the target value is 1, while the probabilities of sampling all other values are 0. The corresponding target distribution can be defined as:
	\[
P_\text{T}(i) = 
	\begin{cases}
1 & \text{if } i = \text{target value} \\
0 & \text{if } i \neq \text{target value}
	\end{cases}
	\]

	Under this setting, the one-hot vector corresponding to the target value in Table~\ref{tab1} is directly used as the probability vector for this state. For example, the one-hot vector corresponding to the target value $0010$ in Table~\ref{tab1} is $[0000000000000100]^\text{T}$. According to the definition of the one-point distribution adopted in this work, this vector indicates that the probability of sampling the state encoded as $0010$ is 1, while the probabilities of sampling all other states are 0. By computing the MSE between the 16-dimensional vector measured after circuit operation and this one-hot vector, the loss value can be obtained.	

	\section{Experimental Results of 4-Qubit Fully Connected Quantum Boltzmann Machine}

	In this work, based on the ‌PennyLane quantum simulation platform, a 4-qubit fully connected QBM is constructed, and computational experiments are conducted. The experiments consist of three parts: first, investigating the effect of parameter $p$ on model convergence by comparing performance across different numbers of QAOA layers; second, evaluating the model's noise robustness under varying noise intensities, critical for its implementation on NISQ devices; third, assessing the model's performance in image generation tasks. All source code, running logs, and experimental data are properly archived in accordance with the reproducibility principle.

	\subsection{Effect of Parameter $p$ on Model Convergence}

\begin{figure}[htbp]
	\centering
	\includegraphics[width=\linewidth]{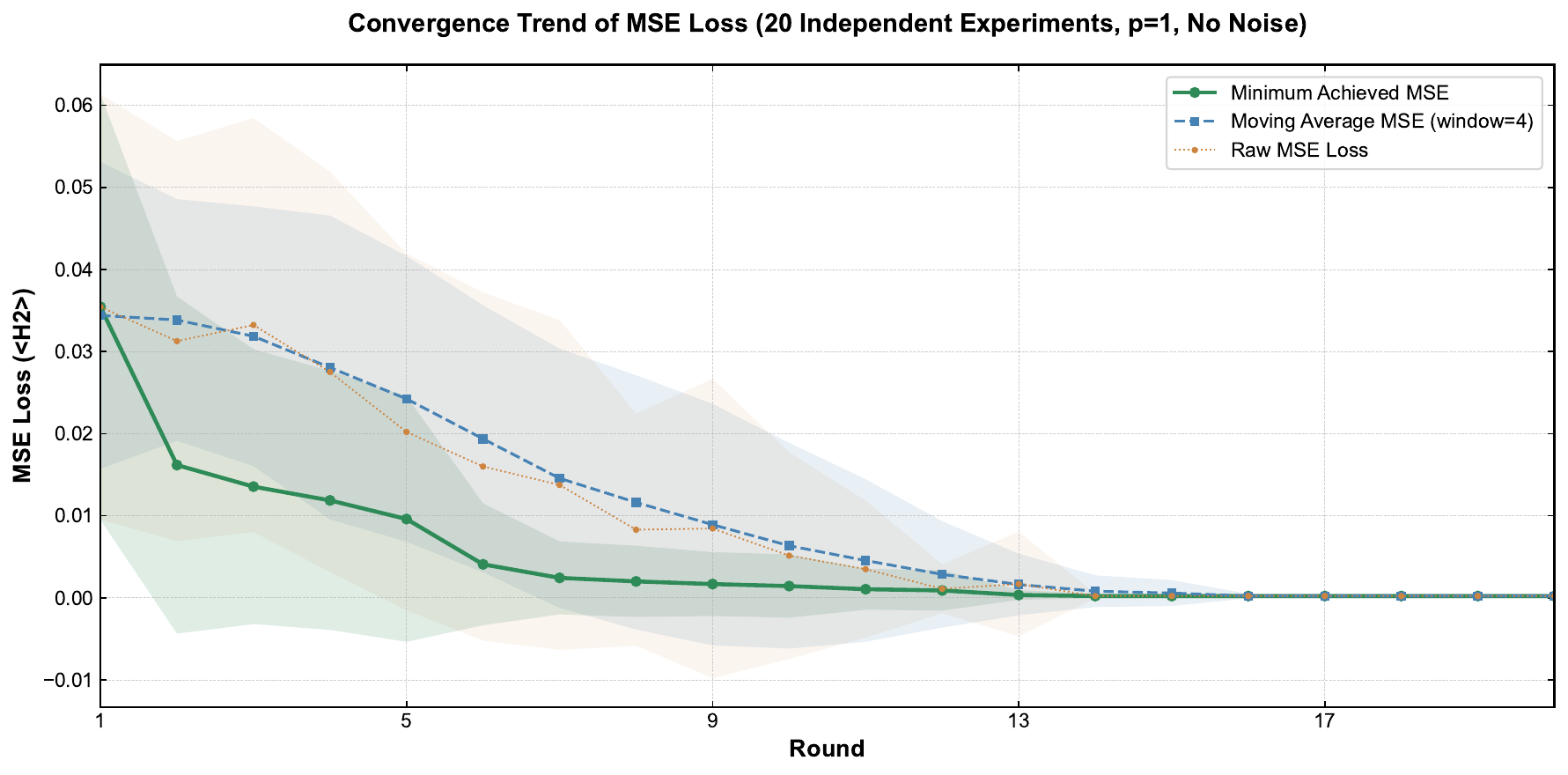}
 	\caption{Experimental results of model convergence computation ($p=1$, no noise)}
	\label{fig3}
	\end{figure}

\begin{figure}[htbp]
	\centering
	\includegraphics[width=\linewidth]{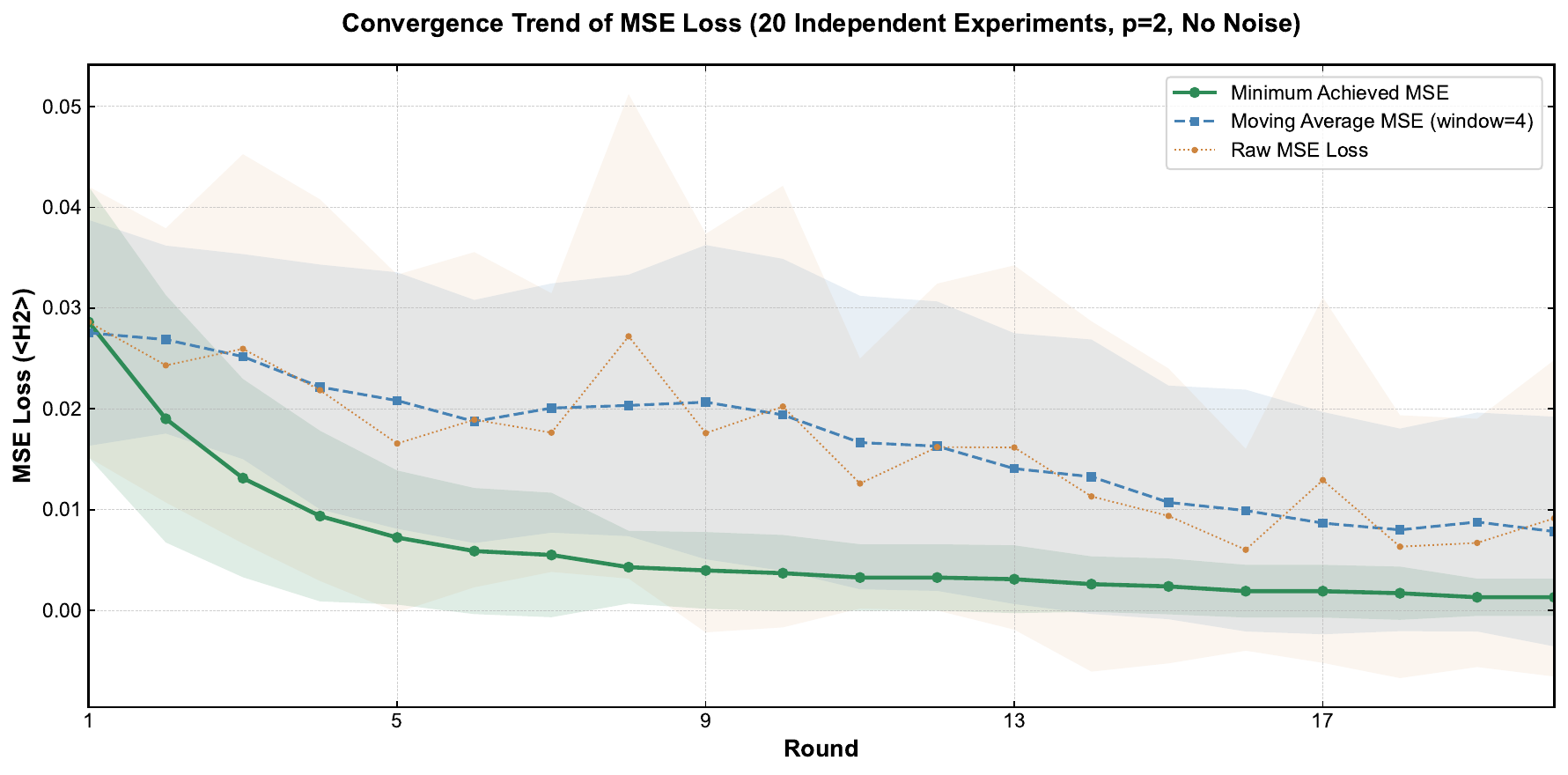}
 	\caption{Experimental results of model convergence computation ($p=2$, no noise)}
	\label{fig4}
	\end{figure}

	This work first investigates the effect of the QAOA layer parameter $p$ on the convergence performance of the proposed QBM. The target distribution to be learned by the model is a one-point distribution centered on the target state $|1001\rangle$, that is, the measurement probability of the quantum state $|1001\rangle$ is 1, while the measurement probabilities of the other 15 quantum states are all 0. Under noiseless conditions, the experiments are divided into two groups corresponding to $p=1$ and $p=2$. To ensure the reliability and accuracy of the experimental results, 20 independent runs are conducted for each group. The experimental results are presented in Figure~\ref{fig3} and Figure~\ref{fig4}, where each figure includes three curves: the raw MSE, the moving average of the MSE, and the minimum achieved MSE. Error bands are plotted based on the standard deviation of the experimental results.

	\clearpage	

	After 20 independent runs, the top 5 quantum states ranked by measurement probability for $p=1$ and $p=2$ under noiseless conditions are shown in Table~\ref{tab2}.

	\clearpage

	\begin{table}[htbp]
  	\centering
  	\caption{Top 5 quantum states by measurement probability (20 independent runs)}
	\label{tab2}
 	\begin{tabular}{c@{\hspace{10pt}}c@{\hspace{10pt}}c@{\hspace{20pt}}c@{\hspace{10pt}}c@{\hspace{10pt}}c}
   \noalign{\hrule height 0.8pt}
   \multicolumn{3}{c}{$p=1$ without noise} & \multicolumn{3}{c}{$p=2$ without noise} \\
   \hline
   States & Mean probabilities & Std. deviations & States & Mean probabilities & Std. deviations \\
   \hline
   $|1001\rangle$ & 0.9559 & 0.0342 & $|1001\rangle$ & 0.9009 & 0.0664 \\
   $|1000\rangle$ & 0.0093 & 0.0088 & $|1101\rangle$ & 0.0174 & 0.0155 \\
   $|1101\rangle$ & 0.0089 & 0.0084 & $|1011\rangle$ & 0.0144 & 0.0128 \\
   $|0001\rangle$ & 0.0088 & 0.0093 & $|0001\rangle$ & 0.0114 & 0.0085 \\
   $|1011\rangle$ & 0.0087 & 0.0087 & $|1000\rangle$ & 0.0101 & 0.0085 \\
   \noalign{\hrule height 0.8pt}
  	\end{tabular}
	\end{table}

	From Figure~\ref{fig3}, Figure~\ref{fig4}, and Table \ref{tab2}, the following conclusions can be drawn: (1) Under noiseless conditions, both fully connected QBMs with $p=1$ and $p=2$ exhibit clear convergence trends, and both are able to approximate the target quantum state with an average probability exceeding $90\%$. (2) By comparison, the model with $p=1$ shows better convergence than that with $p=2$. Specifically, the three curves are more concentrated when $p=1$ and converge to a very low level close to 0 in the later stage; when $p=2$, the three curves exhibit relatively large fluctuations, and their values in the later stage of iteration are significantly higher than those for $p=1$. (3) The average measurement results of the quantum states across 20 independent experiments show that the average probability of measuring the target quantum state $|1001\rangle$ is 0.9559 when $p=1$, which is better than the 0.9009 obtained when $p=2$. Overall, the model performs better when $p=1$. It is speculated that this may be because, as the number of QAOA circuit layers increases, the number of quantum gates increases accordingly, which directly increases the circuit complexity and affects computational accuracy. In view of this, subsequent experiments focus on the noise robustness of the fully connected QBM with $p=1$ under different noise conditions.

	 \subsection{Noise Robustness Experiment}

	In the current NISQ era, noiseless quantum computing devices have not yet been realized, and all commercial quantum computing devices exhibit a certain level of noise. This part of the experiments aims to verify the noise robustness of the fully connected QBM proposed in this work by introducing depolarizing noise into all quantum gates, thereby providing a reference for the subsequent deployment and practical application of the model on real quantum computing devices.

\begin{figure}[htbp]
	\centering
	\includegraphics[width=\linewidth]{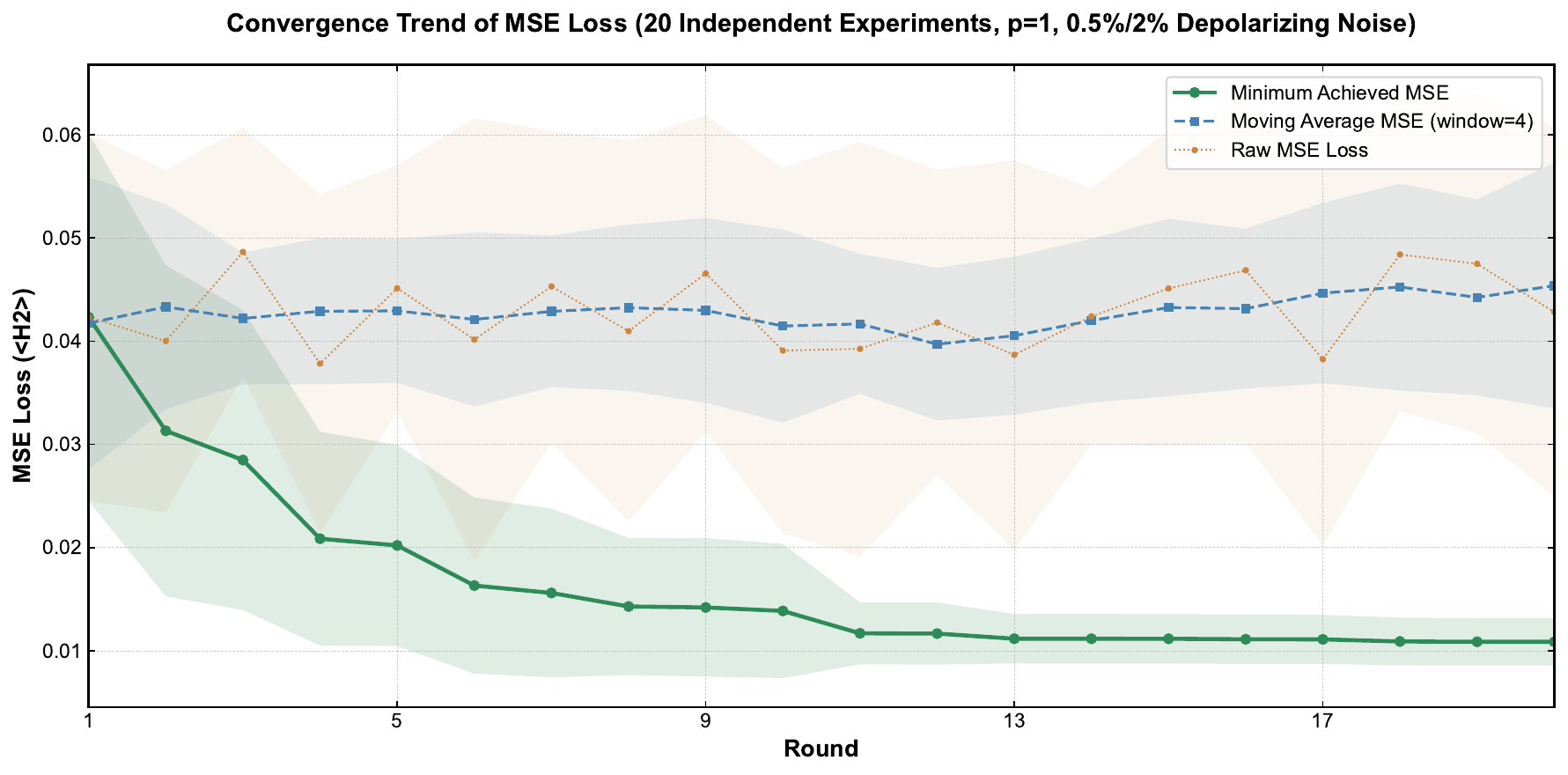}
 	\caption{Experimental results of model convergence computation ($p=1$, single-qubit gate $0.5\%$ and two-qubit gate $2\%$ noise)}
	\label{fig5}
	\end{figure}

\begin{figure}[H]
	\centering
	\includegraphics[width=\linewidth]{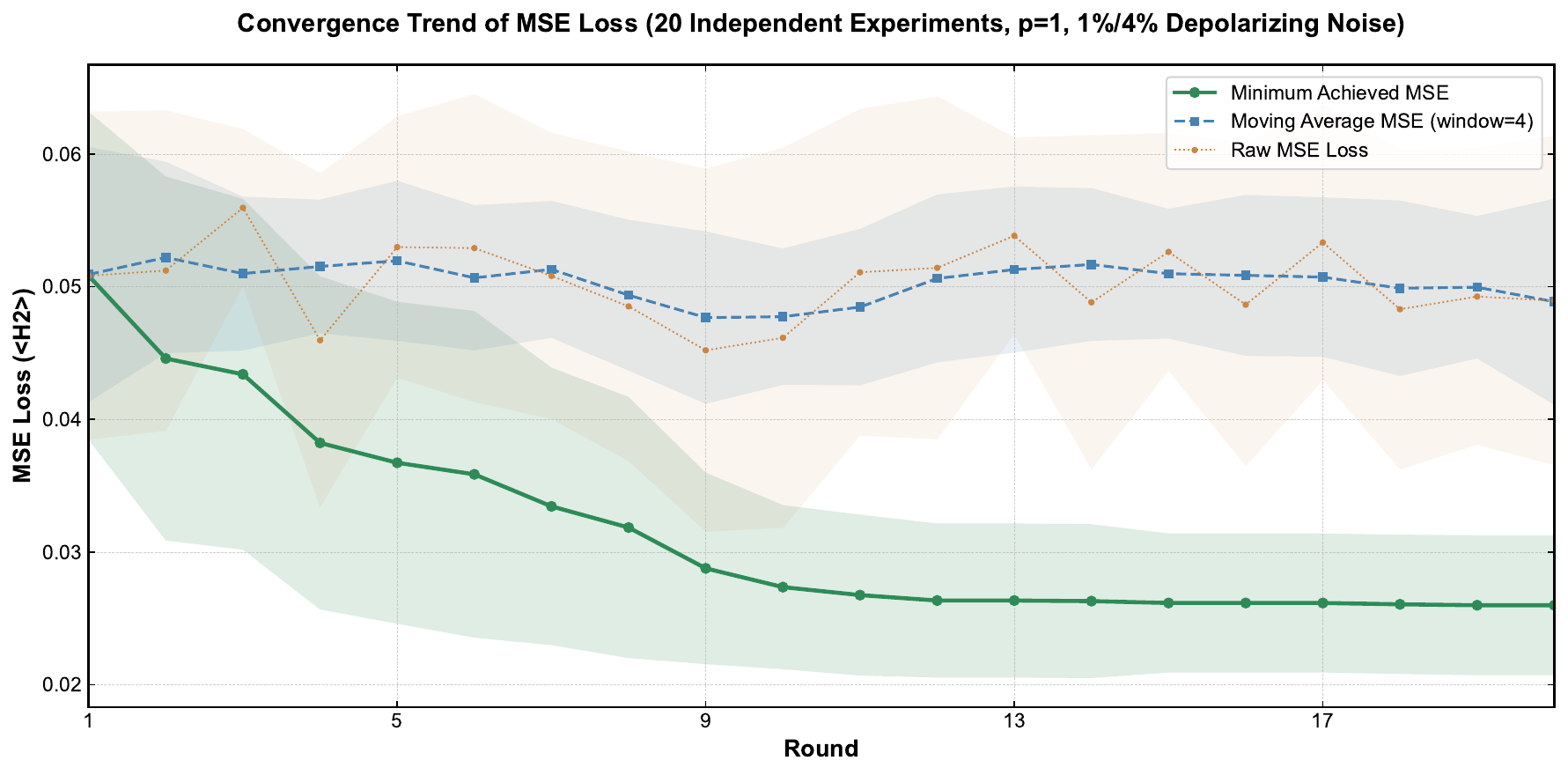}
	\captionsetup{justification=centering}
 	\caption{Experimental results of model convergence computation ($p=1$, single-qubit gate $1\%$ and two-qubit gate $4\%$ noise)}
	\label{fig6}
	\end{figure}

	By querying the public indicators of major quantum computing device manufacturers, it can be seen that as of April 2026 (when the experiments were completed), the typical noise levels of mainstream commercial quantum computing devices are approximately $0.5\%$ for single-qubit gate error rates and $2\%$ for two-qubit gate error rates. Accordingly, to compare the convergence differences of the model under different noise intensities, two sets of noise levels were set under the condition of $p=1$. The first group conducted computational experiments according to the mainstream commercial standards of “single-qubit gate $0.5\%$ and two-qubit gate $2\%$ noise”, while the second group conducted computational experiments under a more stringent noise level with doubled intensity, that is, “single-qubit gate $1\%$ and two-qubit gate $4\%$ noise”. Following the same experimental settings as the previous section, each group was subjected to 20 independent runs. It is worth noting that the noise level of the second group has reached or exceeded the upper noise limit of existing mainstream commercial quantum computing devices. The experimental results are shown in Figure~\ref{fig5} and Figure~\ref{fig6}. Similar to the experiments in the previous section, each figure contains three curves: the raw MSE, the moving average of the MSE, and the minimum achieved MSE, with error bands plotted based on standard deviation values.

	\clearpage

	After 20 independent runs, the top 5 quantum states ranked by measurement probability for $p=1$ under the two noise conditions are shown in Table~\ref{tab3}.

	\clearpage

	\begin{table}[htbp]
  	\centering
  	\caption{Top 5 quantum states by measurement probability (20 independent runs)}
  	\label{tab3}
  	\begin{tabular*}{\linewidth}{@{\extracolsep{\fill}} cccc cc @{}}
    \noalign{\hrule height 0.8pt}
    \multicolumn{3}{c}{$p=1$ with single-qubit gate $0.5\%$} & 
    \multicolumn{3}{c}{$p=1$ with single-qubit gate $1\%$} \\
    \multicolumn{3}{c}{and two-qubit gate $2\%$ noise} & 
    \multicolumn{3}{c}{and two-qubit gate $4\%$ noise} \\
    \hline
    States & Mean probabilities & Std. deviations & States & Mean probabilities & Std. deviations \\
    \hline
    $|1001\rangle$ & 0.6047 & 0.0418 & $|1001\rangle$ & 0.3859 & 0.0537 \\
    $|0001\rangle$ & 0.0703 & 0.0116 & $|1011\rangle$ & 0.0905 & 0.0138 \\
    $|1101\rangle$ & 0.0693 & 0.0112 & $|0001\rangle$ & 0.0880 & 0.0121 \\
    $|1011\rangle$ & 0.0663 & 0.0137 & $|1000\rangle$ & 0.0867 & 0.0156 \\
    $|1000\rangle$ & 0.0618 & 0.0077 & $|1101\rangle$ & 0.0865 & 0.0150 \\
    \noalign{\hrule height 0.8pt}
  	\end{tabular*}
	\end{table}

	From Figure \ref{fig5}, Figure \ref{fig6}, and Table \ref{tab3}, the following conclusions can be drawn: (1) Under the two noise conditions of “single-qubit gate $0.5\%$ and two-qubit gate $2\%$ noise” and “single-qubit gate $1\%$ and two-qubit gate $4\%$ noise”, the model with $p=1$ exhibits good convergence performance. It can approximate the target quantum state with a maximum probability exceeding $38\%$, and the maximum probability is several times higher than that of the second-ranked state, indicating that the model has strong noise robustness. (2) In terms of both the convergence speed and the steady-state value of the minimum achieved MSE, the convergence of the model under “single-qubit gate $0.5\%$ and two-qubit gate $2\%$ noise” is better than that under “single-qubit gate $1\%$ and two-qubit gate $4\%$ noise”, indicating that the increase in noise intensity has a negative impact on model convergence, which is consistent with theoretical expectations. (3) The quantum state measurement results of 20 independent experiments show that the average probability of measuring the target quantum state $|1001\rangle$ under “single-qubit gate $0.5\%$ and two-qubit gate $2\%$ noise” is 0.6047, which is better than 0.3859 under “single-qubit gate $1\%$ and two-qubit gate $4\%$ noise”. It can be seen that even when the noise level reaches or exceeds the upper limit of mainstream commercial quantum computing devices in the NISQ era, the model still shows strong robustness, providing solid experimental support for the subsequent deployment of the model to real quantum devices and practical applications.
	\clearpage	

	\subsection{Image Generation Experiment}

\begin{figure}[htbp]
	\centering
	\includegraphics[width=0.6\linewidth]{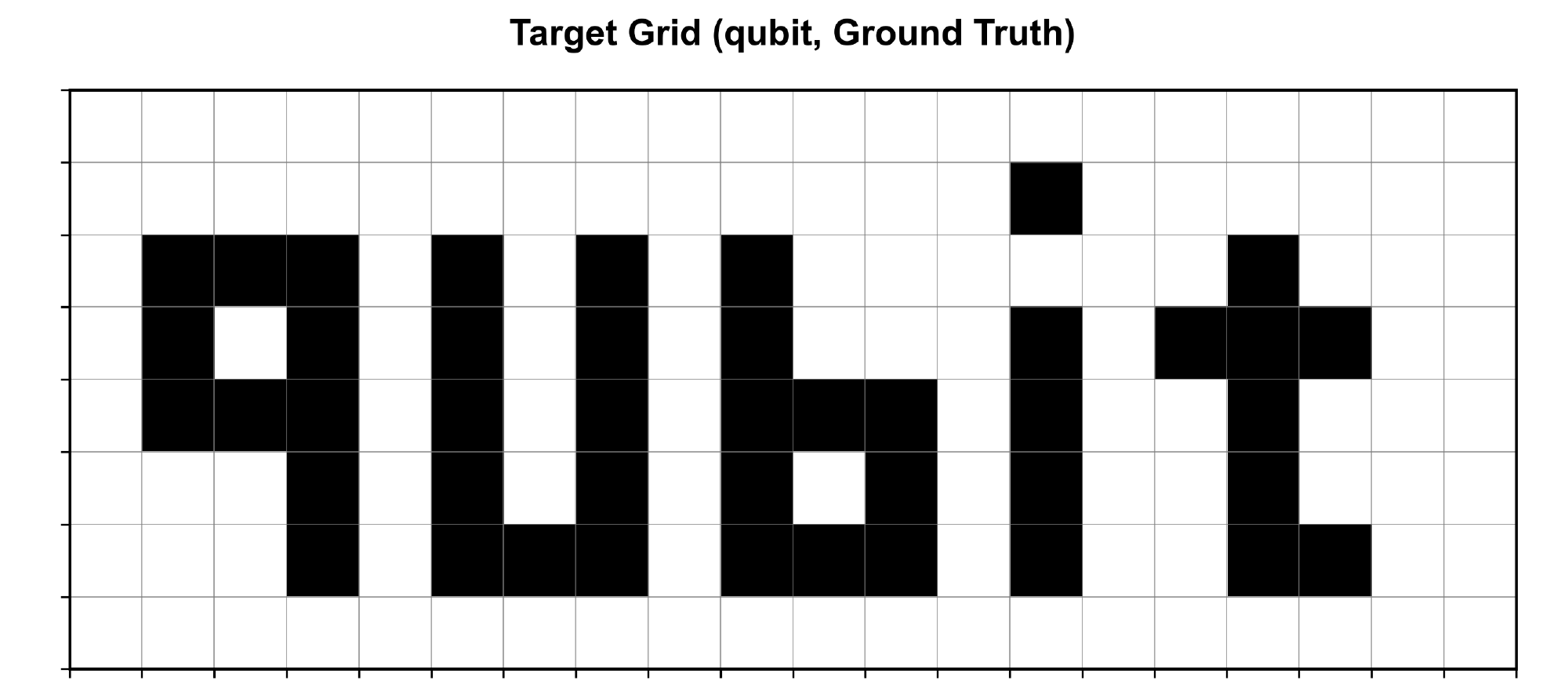}
 	\caption{Grid image of the target distribution}
	\label{fig7}
	\end{figure}

	One of the core tasks of the fully connected QBM is to generate an optimal distribution approximating the target distribution. To this end, this work further conducts image generation experiments based on the proposed model. The target distribution is intuitively visualized by the image of the word “qubit” arranged in an $8\times20$ 2-dimensional grid, as shown in Figure~\ref{fig7}.

	In Figure~\ref{fig7}, each grid cell takes a value of 0 or 1: a value of 0 corresponds to a white-filled cell, while a value of 1 corresponds to a black-filled cell. As defined, the grid consists of $8\times20 = 160$ cells, which can be mapped into a 160-dimensional binary vector with $2^{160}$ possible combinations. The word “qubit” presented in the image is essentially a one-point distribution, meaning that among all possible combinations, the specific combination corresponding to the image is sampled with a probability of 1. The purpose of this experiment is to adopt the fully connected QBM to approximate this one-point distribution, thereby generating an image identical to the target pattern in Figure~\ref{fig7}.

	In this experiment, to further test the image generation capability of the model, two groups of $2 \times 2$ image generation kernels consistent with those in previous experiments are constructed under the condition of $p=1$: one noise-free case and one with typical noise (single-qubit gate $0.5\%$ and two-qubit gate $2\%$ noise). The target distribution in Figure~\ref{fig7} is learned block by block in a left-to-right, top-to-bottom scanning order. Specifically, the overall distribution is divided into $160/4 = 40$ local one-point distributions, which are learned sequentially by the generation kernels. After training, each group of kernels yields 40 sets of optimal parameter configurations. By applying these parameters to the generation tasks of the 40 local image blocks, the complete $8 \times 20$ target grid image can be reconstructed. In accordance with‌ the traceability principle, the source code and the $2 \times 40 = 80$ sets of optimal parameters are properly archived.

\begin{figure}[htbp]
	\centering
	\includegraphics[width=\linewidth]{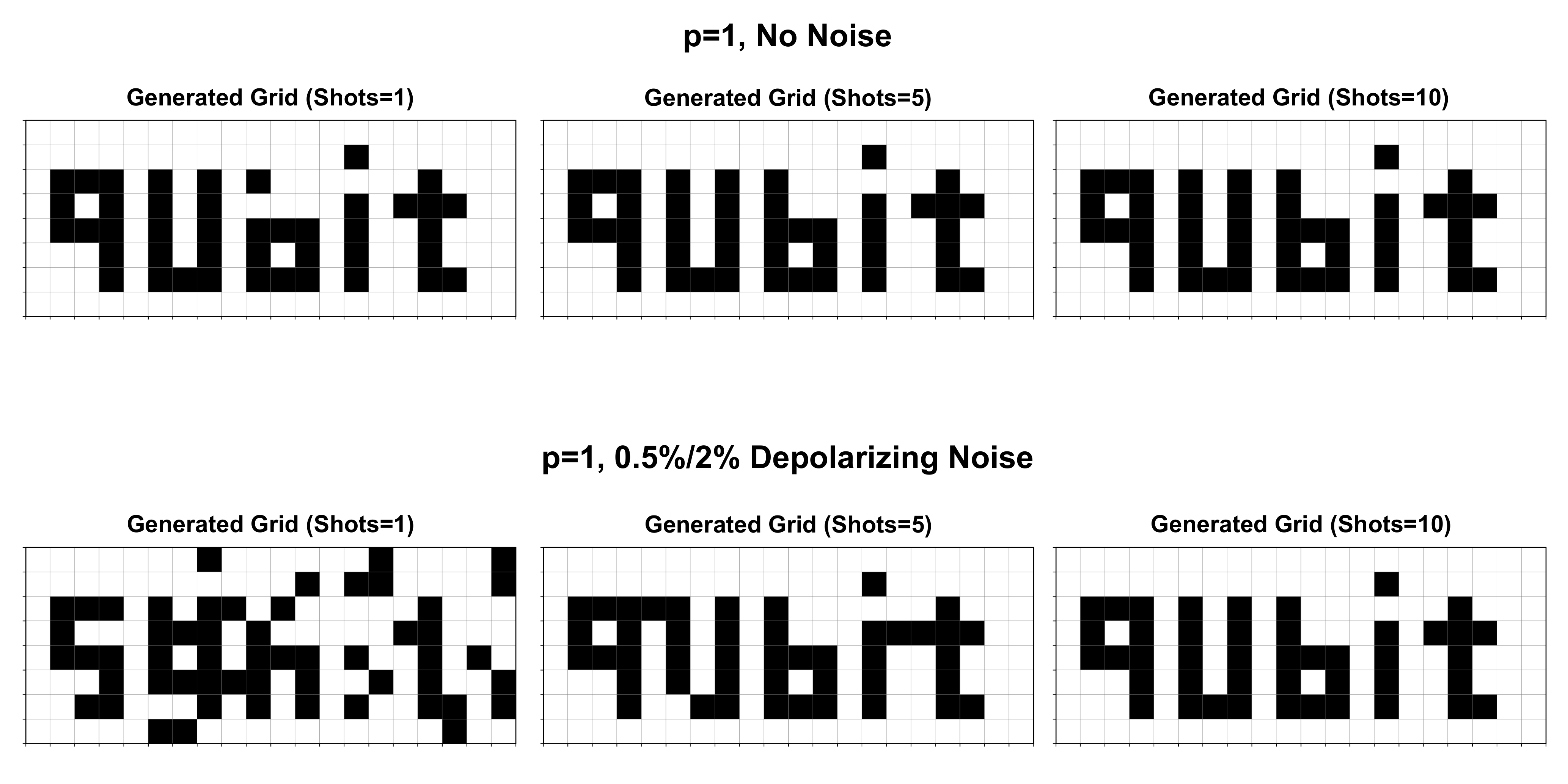}
 	\caption{Results of the image generation experiment}
	\label{fig8}
	\end{figure}

	Image generation is the inverse process of image learning. Using the optimal parameters corresponding to the generation kernels, images are reconstructed block by block. Following the same left-to-right, top-to-bottom order as the learning process, the distribution with the highest measurement probability is taken as the final generation result to build the complete image block by block. Figure~\ref{fig8} intuitively illustrates how the noiseless and noisy generation kernels progressively reconstruct the target image as the number of measurements increases. It can be observed that when the number of measurements is only 10, the word “qubit” matching the target image can be generated regardless of noise interference, indicating that the fully connected QBM proposed in this work has strong anti-interference ability in image generation.

	\section{Conclusion}

	By extending the conventional QAOA circuit to a bilevel optimization structure with optimizable Hamiltonian parameters, this work proposes a fully connected QBM to overcome the limitations of classical partially connected Boltzmann machines and existing mainstream QBMs. The main conclusions are summarized as follows: (1) The model achieves outstanding performance even with only $p=1$, with the average probability of measuring the target quantum state reaching 0.9559 under noiseless conditions. (2) The model also exhibits strong noise robustness. It yields an average measurement probability of 0.6047 under typical noise levels of mainstream commercial quantum computing devices, and 0.3859 at a more stringent noise level with doubled intensity. Both values rank highest among all quantum states and are several times greater than those of the second-ranked state. (3) Combined with a block-by-block learning strategy, the model can reconstruct the target “qubit” image with merely $p=1$ and 10 measurements, regardless of noise disturbance. This superior performance demonstrates its excellent anti-interference capability for image generation tasks.

	\bibliographystyle{unsrtnat}
	\bibliography{references}

\begin{thebibliography}{24}
\providecommand{\natexlab}[1]{#1}
\providecommand{\url}[1]{\texttt{#1}}
\expandafter\ifx\csname urlstyle\endcsname\relax
  \providecommand{\doi}[1]{doi: #1}\else
  \providecommand{\doi}{doi: \begingroup \urlstyle{rm}\Url}\fi

\bibitem[Ackley et~al.(1985)Ackley, Hinton, and Sejnowski]{ackley1985learning}
D.~H. Ackley, G.~E. Hinton, and T.~J. Sejnowski.
\newblock A learning algorithm for {Boltzmann} machines.
\newblock \emph{Cognitive Science}, 9\penalty0 (1):\penalty0 147--169, 1985.
\newblock \doi{10.1207/s15516709cog0901_7}.

\bibitem[Le~Roux and Bengio(2008)]{leroux2008representational}
N.~Le~Roux and Y.~Bengio.
\newblock Representational power of restricted {Boltzmann} machines and deep
  belief networks.
\newblock \emph{Neural Computation}, 20\penalty0 (6):\penalty0 1631--1649,
  2008.
\newblock \doi{10.1162/neco.2008.04-07-510}.

\bibitem[Tieleman(2008)]{tieleman2008training}
T.~Tieleman.
\newblock Training restricted {Boltzmann} machines using approximations to the
  likelihood gradient.
\newblock In \emph{International Conference on Machine Learning (ICML)}, pages
  1064--1071, 2008.

\bibitem[Smolensky(1986)]{smolensky1986information}
P.~Smolensky.
\newblock Information processing in dynamical systems: foundations of harmony
  theory.
\newblock In \emph{Parallel Distributed Processing: Explorations in the
  Microstructure of Cognition, Vol. 1: Foundations}, pages 194--281. MIT Press,
  1986.

\bibitem[Hinton(2002)]{hinton2002training}
G.~E. Hinton.
\newblock Training products of experts by minimizing contrastive divergence.
\newblock \emph{Neural Computation}, 14\penalty0 (8):\penalty0 1771--1800,
  2002.
\newblock \doi{10.1162/089976602760128018}.

\bibitem[Hinton et~al.(2006)Hinton, Osindero, and Teh]{hinton2006fast}
G.~E. Hinton, S.~Osindero, and Y.~W. Teh.
\newblock A fast learning algorithm for deep belief nets.
\newblock \emph{Neural Computation}, 18\penalty0 (7):\penalty0 1527--1554,
  2006.
\newblock \doi{10.1162/neco.2006.18.7.1527}.

\bibitem[Salakhutdinov and Hinton(2009)]{salakhutdinov2009deep}
R.~Salakhutdinov and G.~E. Hinton.
\newblock Deep {Boltzmann} machines.
\newblock In \emph{Proceedings of the International Conference on Artificial
  Intelligence and Statistics (AISTATS)}, pages 448--455, 2009.

\bibitem[Salakhutdinov and Hinton(2012)]{salakhutdinov2012efficient}
R.~Salakhutdinov and G.~E. Hinton.
\newblock An efficient learning procedure for deep {Boltzmann} machines.
\newblock \emph{Neural Computation}, 24\penalty0 (8):\penalty0 1967--2006,
  2012.
\newblock \doi{10.1162/NECO_a_00311}.

\bibitem[Mont{\'u}far et~al.(2014)]{montufar2014number}
G.~Mont{\'u}far et~al.
\newblock On the number of linear regions of deep neural networks.
\newblock In \emph{Advances in Neural Information Processing Systems
  (NeurIPS)}, pages 2924--2932, 2014.

\bibitem[Deng et~al.(2025)]{deng2025interaction}
H.~Deng et~al.
\newblock The interaction bottleneck of deep neural networks: discovery, proof,
  and modulation, 2025.
\newblock URL \url{https://arxiv.org/abs/2512.18607}.

\bibitem[Hinton(2005)]{hinton2005what}
G.~E. Hinton.
\newblock What kind of a graphical model is the brain?
\newblock In \emph{International Joint Conference on Artificial Intelligence
  (IJCAI)}, pages 1765--1775, 2005.

\bibitem[Koller and Friedman(2009)]{koller2009probabilistic}
D.~Koller and N.~Friedman.
\newblock \emph{Probabilistic Graphical Models: Principles and Techniques}.
\newblock MIT Press, 2009.
\newblock ISBN 9780262013192.

\bibitem[Wiebe et~al.(2014)]{wiebe2014quantum}
N.~Wiebe et~al.
\newblock Quantum deep learning.
\newblock 2014.
\newblock URL \url{https://arxiv.org/abs/1412.3489}.

\bibitem[Wiebe and Wossnig(2019)]{wiebe2019generative}
N.~Wiebe and L.~Wossnig.
\newblock Generative training of quantum {Boltzmann} machines with hidden
  units, 2019.
\newblock URL \url{https://arxiv.org/abs/1905.09902}.

\bibitem[Srivastava and Sundararaghavan(2023)]{srivastava2023generative}
S.~Srivastava and V.~Sundararaghavan.
\newblock Generative and discriminative training of {Boltzmann} machine through
  quantum annealing.
\newblock \emph{Scientific Reports}, 13\penalty0 (1):\penalty0 12456, 2023.
\newblock \doi{10.1038/s41598-023-34652-4}.

\bibitem[Zoufal et~al.(2020)Zoufal, Lucchi, and Woerner]{zoufal2020variational}
C.~Zoufal, A.~Lucchi, and S.~Woerner.
\newblock Variational quantum {Boltzmann} machines, 2020.
\newblock URL \url{https://arxiv.org/abs/2006.06004}.

\bibitem[Coopmans and Benedetti(2024)]{coopmans2024sample}
L.~Coopmans and M.~Benedetti.
\newblock On the sample complexity of quantum {Boltzmann} machine learning.
\newblock \emph{Communications Physics}, 7:\penalty0 274, 2024.
\newblock \doi{10.1038/s42005-024-01763-x}.

\bibitem[Patel et~al.(2024)]{piatkowski2024quantum}
D.~Patel et~al.
\newblock Quantum {Boltzmann} machine learning of ground-state energies, 2024.
\newblock URL \url{https://arxiv.org/abs/2410.12935}.

\bibitem[Demidik et~al.(2025)]{demidik2025expressive}
M.~Demidik et~al.
\newblock Expressive equivalence of classical and quantum restricted
  {Boltzmann} machines, 2025.
\newblock URL \url{https://arxiv.org/abs/2502.17562}.

\bibitem[Kimura et~al.(2025)Kimura, Kato, and Hayashi]{kimura2025structured}
T.~Kimura, K.~Kato, and M.~Hayashi.
\newblock Structured quantum learning via em algorithm for {Boltzmann} machines
  ({Note}: The lowercase “em” in the title is as per the original paper;
  standard abbreviation is {EM} (expectation-maximization).), 2025.
\newblock URL \url{https://arxiv.org/abs/2507.21569}.

\bibitem[Rule and Rrapaj(2025)]{rule2025exact}
E.~Rule and E.~Rrapaj.
\newblock Exact block encoding of imaginary time evolution with universal
  quantum neural networks.
\newblock \emph{Physical Review Research}, 7\penalty0 (1):\penalty0 013306,
  2025.
\newblock \doi{10.1103/PhysRevResearch.7.013306}.

\bibitem[Farhi et~al.(2000)]{farhi2000quantum}
E.~Farhi et~al.
\newblock Quantum computation by adiabatic evolution, 2000.
\newblock URL \url{https://arxiv.org/abs/quant-ph/0001106}.

\bibitem[Farhi et~al.(2014)Farhi, Goldstone, and Gutmann]{farhi2014quantum}
E.~Farhi, J.~Goldstone, and S.~Gutmann.
\newblock A quantum approximate optimization algorithm, 2014.
\newblock URL \url{https://arxiv.org/abs/1411.4028}.

\bibitem[Mitarai et~al.(2018)]{mitarai2018quantum}
K.~Mitarai et~al.
\newblock Quantum circuit learning.
\newblock \emph{Physical Review A}, 98\penalty0 (3):\penalty0 032309, 2018.
\newblock \doi{10.1103/PhysRevA.98.032309}.

\end{thebibliography}
	
\end{document}